%% file: main.tex
\documentclass[10pt,journal,compsoc]{IEEEtran}
\ifCLASSOPTIONcompsoc
  \usepackage[nocompress]{cite}
\else
  \usepackage{cite}
\fi
\ifCLASSINFOpdf
\else
\fi

\usepackage{adjustbox}
\usepackage{graphicx,amssymb,lineno}
\usepackage{algorithm}
\usepackage{algorithmic}
\usepackage{multirow}
\usepackage{epsfig}
\usepackage{amssymb}
\usepackage{footmisc}
\usepackage{footnpag}
\usepackage{subfigure}
\usepackage{amsmath,paralist}
\usepackage{color}
\usepackage{perpage}
\usepackage{epstopdf}
\usepackage{color,soul}
\usepackage{textcomp}
\usepackage{listings} 

\begin{document}
%
\title{SZ3: A Modular Framework for Composing Prediction-Based Error-Bounded Lossy Compressors}
%
%
%
%

\author{Xin~Liang\IEEEauthorrefmark{1},~
        Kai~Zhao\IEEEauthorrefmark{1},~
        Sheng~Di,~\IEEEmembership{Senior Member,~IEEE,}
        Sihuan~Li,~
        Robert~Underwood,~\\
        Ali~M.~Gok,~
        Jiannan~Tian,~
        Junjing~Deng,~
        Jon~C.~Calhoun,~
        Dingwen~Tao,~\\
        Zizhong~Chen,~\IEEEmembership{Senior Member,~IEEE,}
        and~Franck~Cappello,~\IEEEmembership{Fellow,~IEEE}
\IEEEcompsocitemizethanks{
\IEEEcompsocthanksitem X. Liang is with the department of Computer Science, Missouri University of Science and Technology, Rolla,
MO 65409.\protect\\
E-mail: xliang@mst.edu
\IEEEcompsocthanksitem K. Zhao and Z. Chen are with the Department
of Computer Science \& Engineering, University of California, Riverside, Riverside,
CA 92521.\protect\\
E-mail: kzhao016@ucr.edu, chen@cs.ucr.edu
\IEEEcompsocthanksitem S. Li is with Facebook Inc., Menlo Park, CA 94025.\protect\\
E-mail: sli049@ucr.edu
\IEEEcompsocthanksitem S. Di and F. Cappello are with the Mathematics and Computer Science Division, Argonne National Laboratory, Lemont, IL 60439.\protect\\
E-mail: sdi1@anl.gov, cappello@mcs.anl.gov
\IEEEcompsocthanksitem J. Deng is with the Advanced Photon Source, Argonne National Laboratory, Lemont, IL 60439.\protect\\
E-mail: junjingdeng@anl.gov
\IEEEcompsocthanksitem A. M. Gok is with Cerebras Systems, Sunnyvale, CA 94085.\protect\\
E-mail: ali.gok@cerebras.net
\IEEEcompsocthanksitem J. Tian and D. Tao are with the Department
of Computer Science, Washington State University, Pullman, WA 99164.\protect\\
E-mail: \{jiannan.tian, dingwen.tao\}@wsu.edu
\IEEEcompsocthanksitem R. Underwood is with the School of Computing, Clemson University, Clemson, SC 29634.\protect\\
E-mail: robertu@g.clemson.edu
\IEEEcompsocthanksitem J.C. Calhoun is with the Holcombe Department of Electrical and Computer Engineering, Clemson University, Clemson, SC 29634.\protect\\
E-mail: jonccal@clemson.edu 
\IEEEcompsocthanksitem \IEEEauthorrefmark{1} The two authors contributed equally to this paper.\protect
\IEEEcompsocthanksitem Sheng Di is the corresponding author.
\protect\\

}

}

%
%

\markboth{Journal of \LaTeX\ Class Files,~Vol.~14, No.~8, August~2015}%
{Shell \MakeLowercase{\textit{et al.}}: Bare Demo of IEEEtran.cls for Computer Society Journals}
%



\IEEEtitleabstractindextext{%
\input{tex/abstract}

\begin{IEEEkeywords}
Big Data, Error-Bounded Lossy Compression, Data Reduction, Large-Scale Scientific Simulation
\end{IEEEkeywords}}

\maketitle

\IEEEdisplaynontitleabstractindextext

%
\IEEEpeerreviewmaketitle

\newcommand\xin[1]{{\color{cyan}{Xin: #1}}}
\newcommand\robert[1]{{\color{green}{Robert: #1}}}

\input{tex/introduction}
\input{tex/related}

\input{tex/design}
\input{tex/usecases}
\input{tex/conclusion}


%

\appendices

\section{}
We demonstrate some representative interfaces and functions in this appendix. 
Note that \texttt{T} is the template for data type, \texttt{N} is the template for dimensionality, and \texttt{X} is the template for quantized data type.
\subsection{Snippet of Preprocess Interface}
\label{appd:preprocess}
\begin{lstlisting}[language=C++,basicstyle=\footnotesize,breaklines=true]
template<class T, uint N>
class PreprocessInterface {
  virtual void preprocess(T * data, SZ::Config<T, N>& conf);
  virtual void postprocess(T * data, SZ::Config<T, N>& conf);
};
\end{lstlisting}

\subsection{Snippet of Predictor Interface}
\label{appd:predictor}
\begin{lstlisting}[language=C++,basicstyle=\footnotesize,breaklines=true]
template<class T, uint N>
class PredictorInterface {
  virtual T predict(const iterator &iter);
  virtual T estimate_error(const iterator &iter);
  virtual uint save(uchar *&c);
  virtual void load(uchar *&c);
};
\end{lstlisting}

\subsection{Snippet of Quantizer Interface}
\label{appd:quantizer}
\begin{lstlisting}[language=C++,basicstyle=\footnotesize,breaklines=true]
template<class T, class X, uint N>
class QuantizerInterface {
  virtual X quantize(T data, T pred);
  virtual T recover(T pred, X quant_value);
  virtual uint save(uchar *&c);
  virtual void load(uchar *&c);
};
\end{lstlisting}

\subsection{Snippet of Encoder Interface}
\label{appd:encoder}
 \begin{lstlisting}[language=C++,basicstyle=\footnotesize,breaklines=true]
 template<class T>
 class EncoderInterface {
  virtual size_t encode(vector<T> &bins, uchar *&bytes);
  virtual vector<T> decode(uchar *&bytes, size_t length);
  virtual uint save(uchar *&c);
  virtual void load(uchar *&c);
};
\end{lstlisting}

\subsection{Snippet of Lossless Interface}
\label{appd:lossless}
 \begin{lstlisting}[language=C++,basicstyle=\footnotesize,breaklines=true]
class LosslessInterface {
    virtual uchar *compress(uchar *data, size_t inSize, size_t &outSize);
     virtual uchar *decompress(uchar *data, size_t& outSize);
};
\end{lstlisting}

\subsection{Snippet of Compressor Class}
\label{appd:compress}
\begin{lstlisting}[language=C++,basicstyle=\footnotesize,breaklines=true]
template<class T, size_t N, class Preprocessor, class Predictor, class Quantizer, class Encoder, class Lossless>
class SZ_Compressor {..}
\end{lstlisting}

\subsection{Snippet of Prediction and Quantization}
\label{appd:pd}
\begin{lstlisting}[language=C++,basicstyle=\footnotesize,breaklines=true]
vector<int> predict_quantize(T *data) {
    multidimensional_iter blocks(data)
    for (auto block = blocks->begin(); block!=blocks->end(); ++block) {
        for (auto element = block->begin(); element != block->end(); ++element) {
            pred=predictor.predict(element);
            quan=quantizer.quantize(*element, pred);
            quantization_results.push_back(quan);
        }
    }
    return quantization_results;
}
\end{lstlisting}

\ifCLASSOPTIONcompsoc
  \section*{Acknowledgments}
\else
  \section*{Acknowledgment}
\fi

This research was supported by the Exascale Computing Project (ECP), Project Number: 17-SC-20-SC, a collaborative effort of two DOE organizations – the Office of Science and the National Nuclear Security Administration, responsible for the planning and preparation of a capable exascale ecosystem, including software, applications, hardware, advanced system engineering and early testbed platforms, to support the nation’s exascale computing imperative. The material was supported by the U.S. Department of Energy (DOE), Office of Science and DOE Advanced Scientific Computing Research (ASCR) office, under contract DE-AC02-06CH11357, and supported by the National Science Foundation under Grant OAC-2003709, OAC-2003624/2042084, SHF-1910197, and OAC-2104023. This research used resources of the Advanced Photon Source, a U.S. Department of Energy (DOE) Office of Science User Facility, operated for the DOE Office of Science by Argonne National Laboratory under Contract No. DE-AC02-06CH11357. We acknowledge the computing resources provided on Bebop, which is operated by the Laboratory Computing Resource Center at Argonne National Laboratory.

\ifCLASSOPTIONcaptionsoff
  \newpage
\fi



%
\bibliographystyle{IEEEtran}
\bibliography{bib/refs}




%






\begin{IEEEbiography}[{\includegraphics[width=1in,height=1.25in,clip,keepaspectratio]{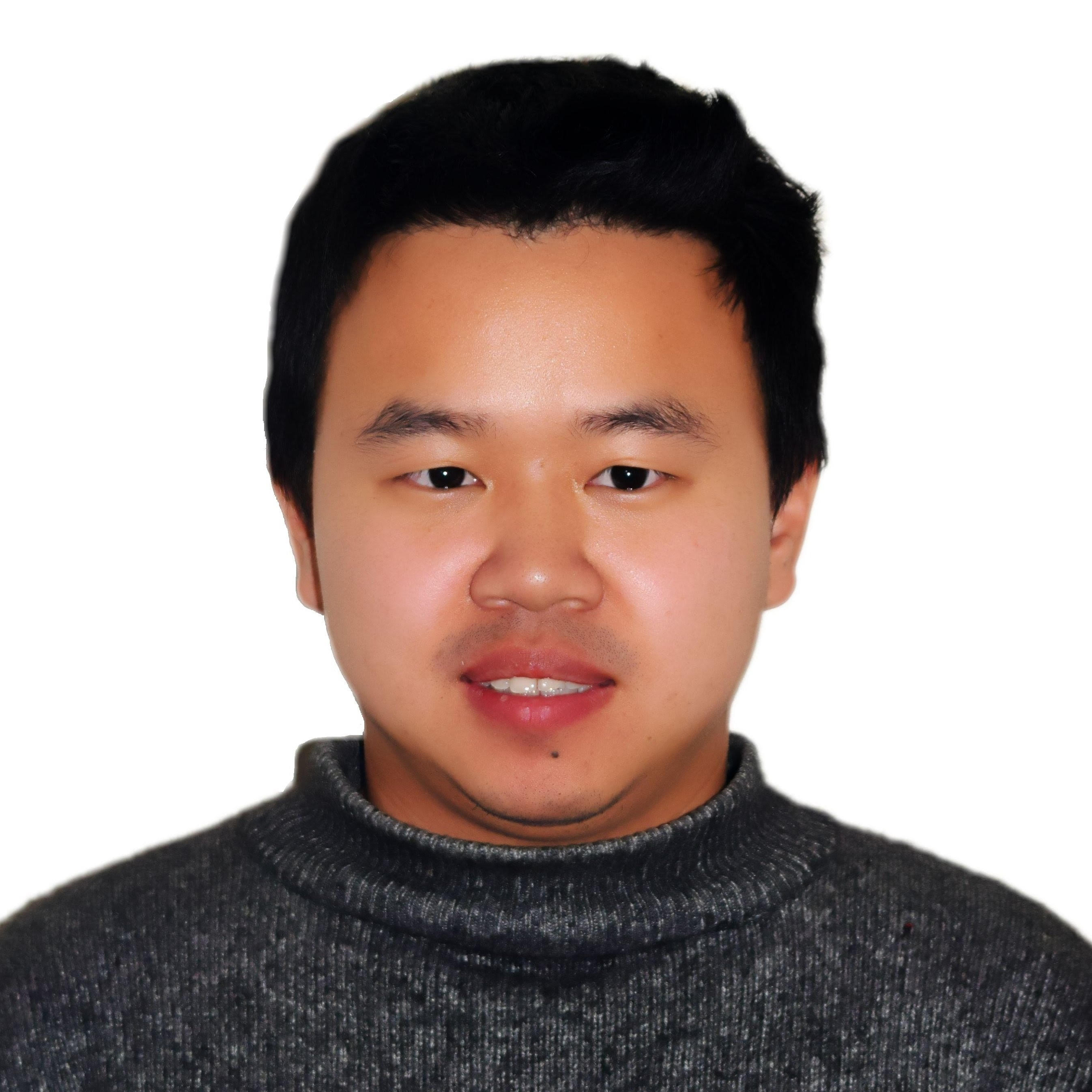}}]{Xin Liang} is an assistant professor with the Department of Computer Science at Missouri University of Science \& Technology. Prior to that, he worked as a Computer/Data Scientist in the Workflow Systems Group at Oak Ridge National Laboratory. He received his Ph.D. degree from University of California, Riverside in 2019 and his bachelor's degree from Peking University in 2014. His research interests include high-performance computing, parallel and distributed systems, scientific data management and reduction, big data analytic, and scientific visualization. He has interned in multiple national laboratories and worked on several exascale computing projects. He is a member of the IEEE. Email: xliang@mst.edu.
\end{IEEEbiography}


\begin{IEEEbiography}[{\includegraphics[width=1in,height=1.25in,clip,keepaspectratio]{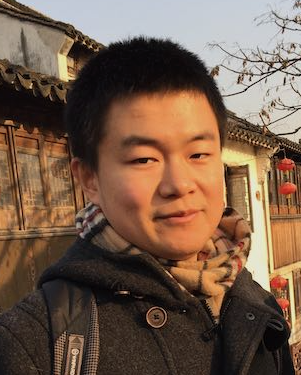}}]{Kai Zhao}
received his bachelor's degree from Peking University in 2014 and will receive his Ph.D. degree from University of California, Riverside in 2022. He is a long-term intern at Argonne National Laboratory.
His research interests include high-performance computing, scientific data management and reduction, and resilient machine learning. Email: kzhao016@ucr.edu.
\end{IEEEbiography}

\vskip -3\baselineskip plus -1fil

\begin{IEEEbiography}[{\includegraphics[width=1in,height=1.25in,clip,keepaspectratio]{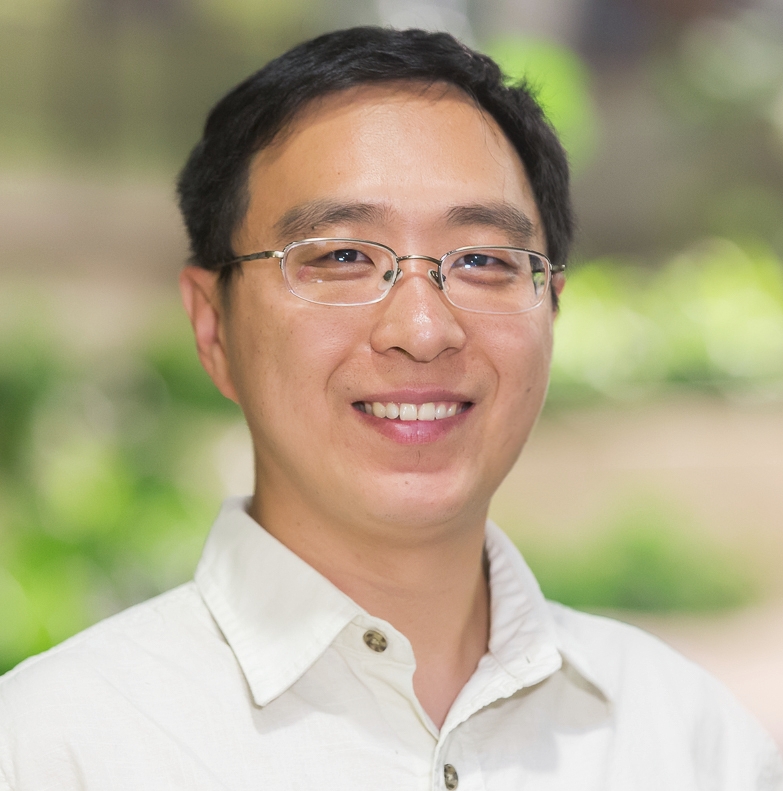}}]{Sheng Di}
(Senior Member, IEEE) received his master's degree from Huazhong University of Science and Technology in 2007 and Ph.D. degree from the University of Hong Kong in 2011. He is currently a computer scientist at Argonne National Laboratory. 
Dr. Di's research interest involves resilience on high-performance computing (such as silent data corruption, optimization checkpoint model, and in-situ data compression) and broad research topics on cloud computing (including optimization of resource allocation, cloud network topology, and prediction of cloud workload/hostload).
He is working on multiple HPC projects, such as detection of silent data corruption, characterization of failures and faults for HPC systems, and optimization of multilevel checkpoint models. He is the recipient of DOE 2021 Early Career Research Program Award. Email: sdi1@anl.gov.
\end{IEEEbiography}

\vskip -2\baselineskip plus -1fil

\begin{IEEEbiography}[{\includegraphics[width=1in,height=1.25in,clip,keepaspectratio]{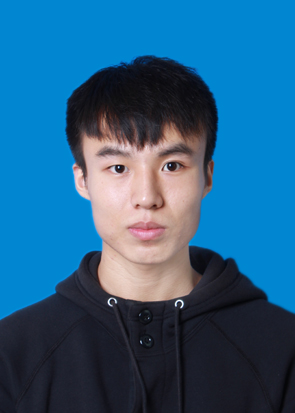}}]{Sihuan Li}
is a research scientist at Facebook. Before that, he received his Ph.D. degree in computer science at University of California, Riverside. He obtained his bachelor's degree in math from Huazhong University of Science and Technology, China. He did a long-term internship at Argonne National Laboratory. Broadly speaking, his research interests fall into High Performance Computing. Specifically, he mainly studies Algorithm Based Fault Tolerance (ABFT), lossy compression and their applications in large scale scientific simulations. Email: sli049@ucr.edu.
\end{IEEEbiography}

\vskip -2\baselineskip plus -1fil

\begin{IEEEbiography}[{\includegraphics[width=1in,height=1.25in,clip,keepaspectratio]{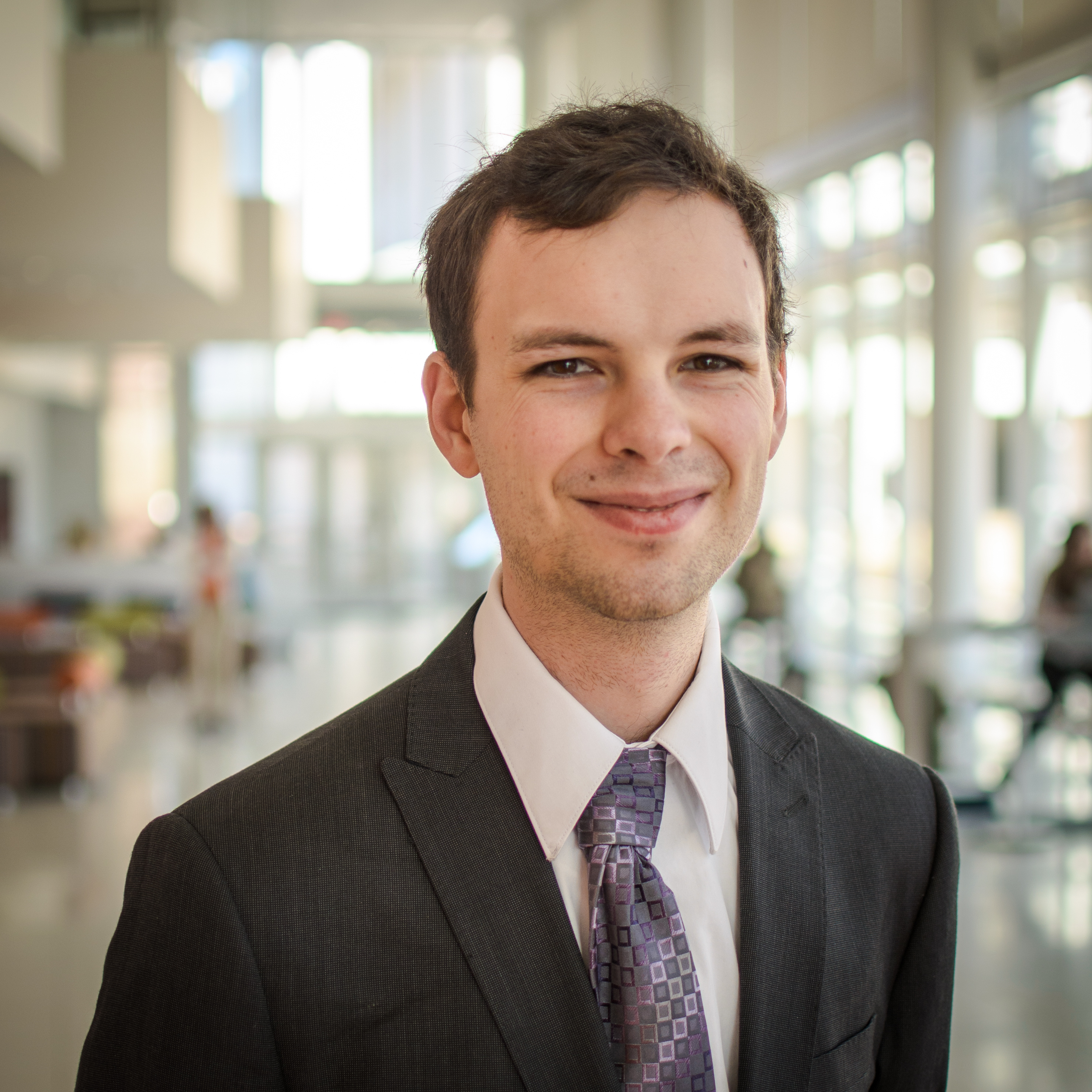}}]{Robert Underwood}
is a PhD Candidate at Clemson University.
His research interests involve using approximate computing methods such as
lossy data compression to accelerate parallel and distributed computing while ensuring that scientific data integrity is preserved.
He is currently working on using optimization based approaches to configure lossy compression.
Email: robertu@g.clemson.edu
\end{IEEEbiography}
\vskip -2\baselineskip plus -1fil

\begin{IEEEbiography}[{\includegraphics[width=1in,height=1.25in,clip,keepaspectratio]{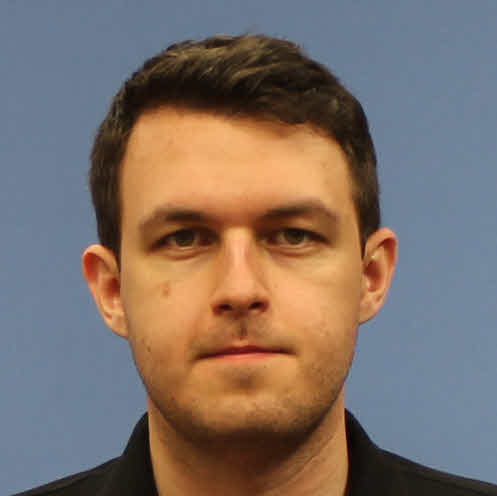}}]{Ali Murat Gok} is a computer engineer at Cerebras Systems. He received his PhD in computer engineering from Northwestern University in 2018 and his B.S. in electronics engineering / mathematics double major program from Bogazici University, Turkey in 2012. He had long term internships followed by a postdoctoral position at Argonne National Laboratory. His research interest include energy efficient computer architecture, high performance computing, and scientific lossy data compression. Email: ali.gok@cerebras.net
\end{IEEEbiography}

\vskip -2\baselineskip plus -1fil

\begin{IEEEbiography}[{\includegraphics[width=1in,height=1.25in,clip,keepaspectratio]{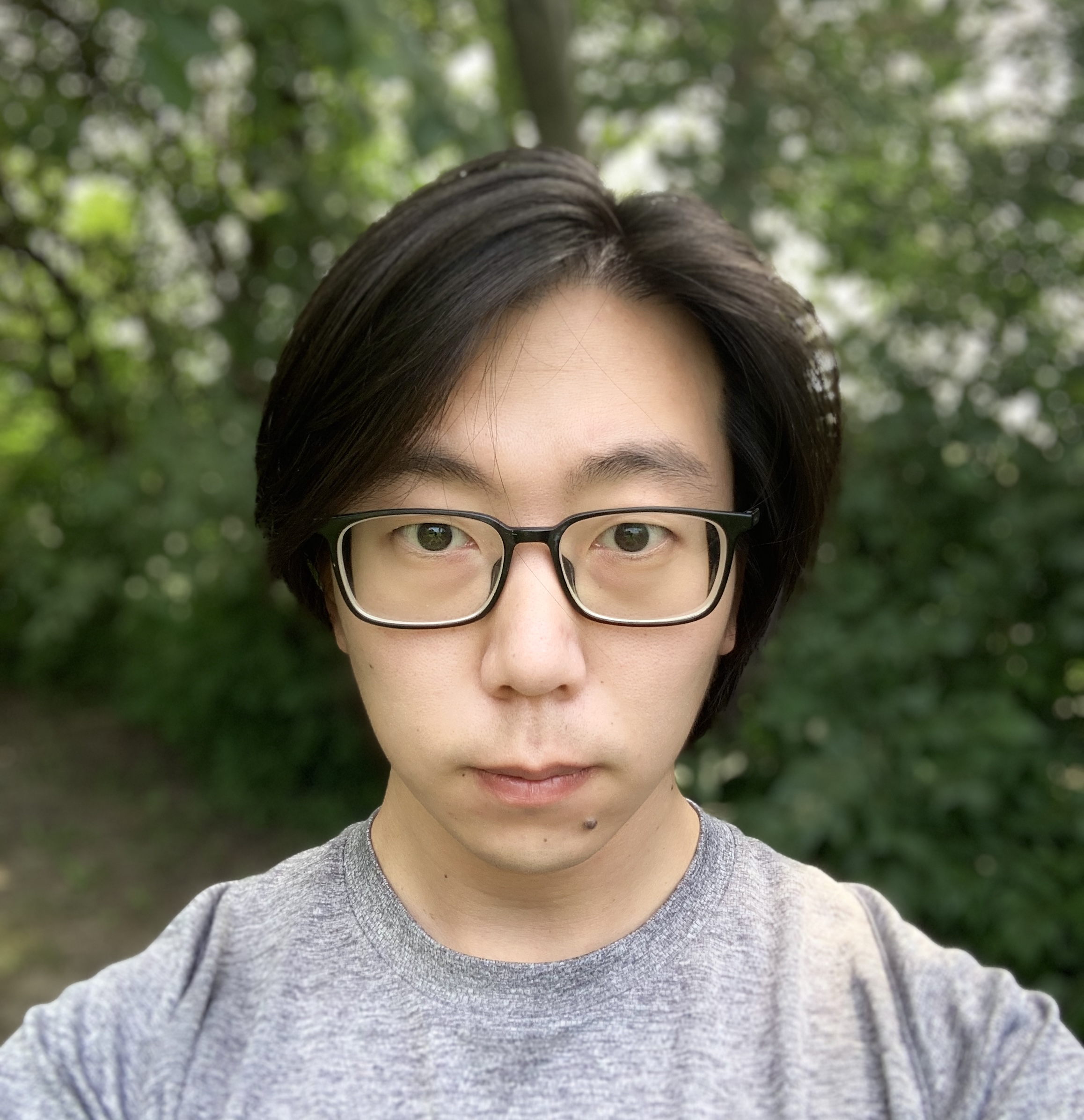}}]{Jiannan Tian}
is current PhD Candidate in Computer Science at Washington State University. His research interests include lossy compression for scientific data and error analysis, and GPU-centric computing.
His ongoing project including developing GPU-accelerated compression algorithm and system design optmization of lossy compression framework.
Email: jiannan.tian@wsu.edu
\end{IEEEbiography}

\vskip -2\baselineskip plus -1fil

\begin{IEEEbiography}[{\includegraphics[width=1in,height=1.25in,clip,keepaspectratio]{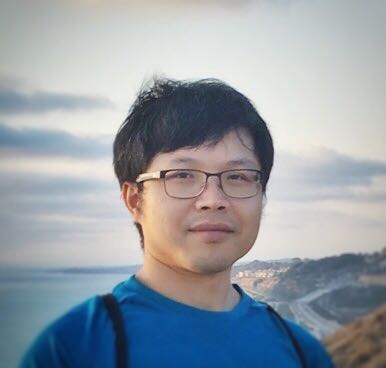}}]{Junjing Deng} 
is a physicist at the Advanced Photon Source, Argonne National Laboratory. He received the Ph.D. degree in applied physics from Northwestern University in 2016. His research interests center on high-resolution synchrotron X-ray microcopy, lensless computational imaging, and their applications on a variety of scientific problems. Email: junjingdeng@anl.gov.
\end{IEEEbiography}

\vskip -2\baselineskip plus -1fil

\begin{IEEEbiography}[{\includegraphics[width=1in,height=1.25in,clip,keepaspectratio]{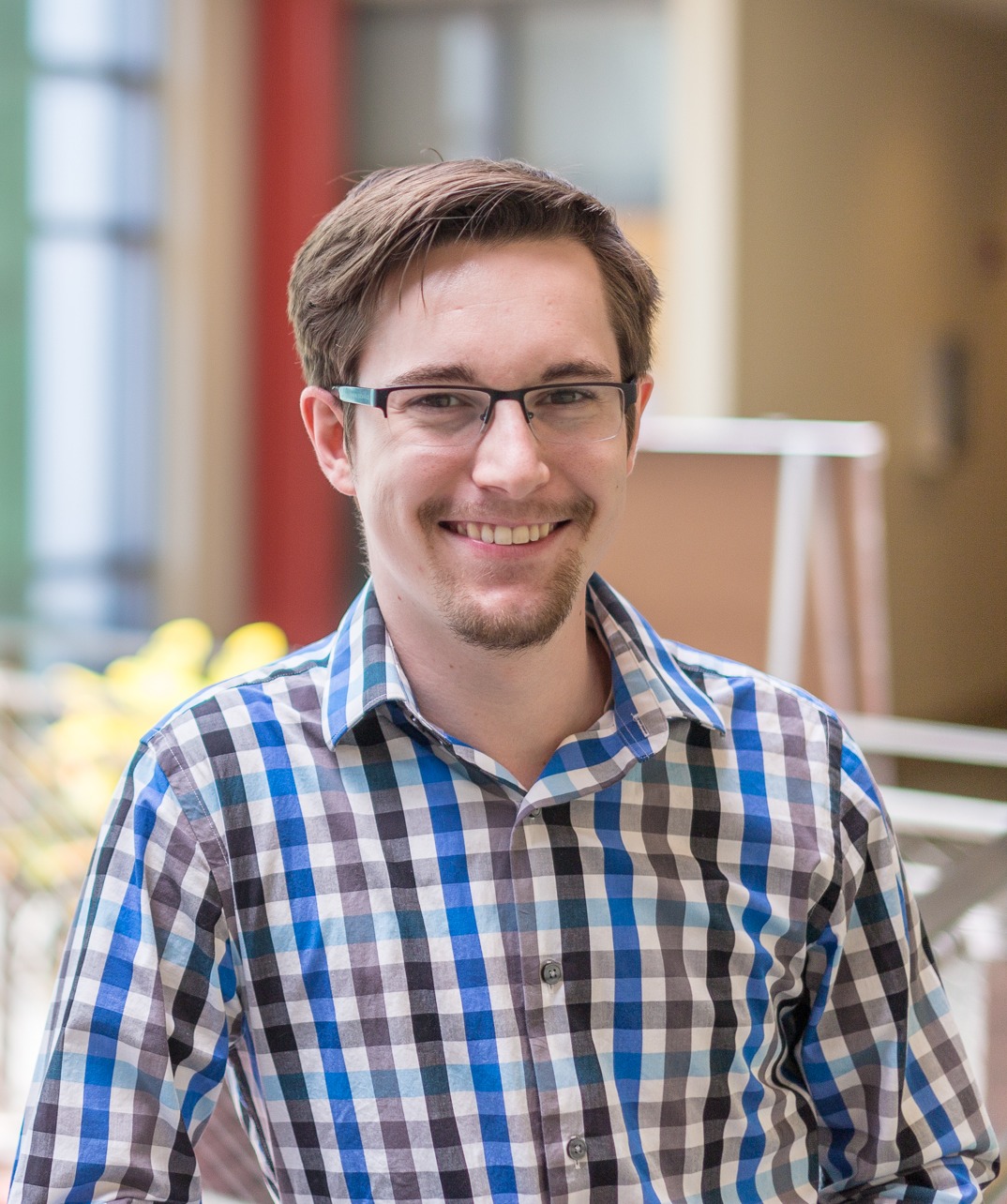}}]{Jon C. Calhoun}
is an Assistant Professor in the Holcombe Department of Electrical and Computer Engineering at Clemson University. He received a B.S. in Computer Science from Arkansas State University in 2012, a B.S. in Mathematics from Arkansas State University in 2012, and a Ph.D. in Computer Science from the University of Illinois at Urbana-Champaign in 2017. His research interests lie in fault tolerance and resilience for high-performance computing (HPC) systems and applications, lossy and lossless data compression, scalable numerical algorithms, power-aware computing, and approximate computing.
\end{IEEEbiography}

\vskip -2\baselineskip plus -1fil

\begin{IEEEbiography}[{\includegraphics[width=1in,height=1.25in,clip,keepaspectratio]{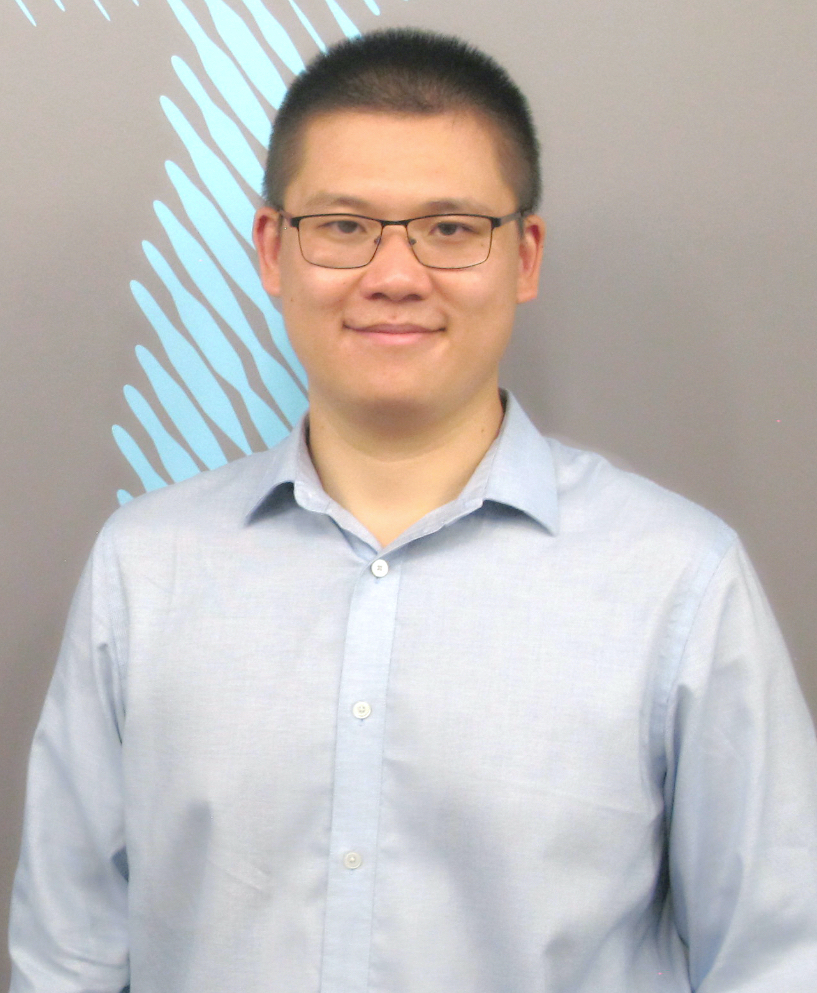}}]{Dingwen Tao}
is an assistant professor of computer science at Washington State University. He received his Ph.D. in Computer Science from University of California, Riverside in 2018 and B.S. in Mathematics from University of Science and Technology of China in 2013. He works at the intersection of HPC and big data analytics, focusing on scientific data management, HPC storage and I/O, fault tolerance at extreme scale, and distributed machine learning. 
He was the receipt of the 2020 IEEE Computer Society TCHPC Early Career Researchers Award for Excellence in HPC.
\end{IEEEbiography}

\vskip -2\baselineskip plus -1fil

\begin{IEEEbiography}[{\includegraphics[width=1in,height=1.25in,clip,keepaspectratio]{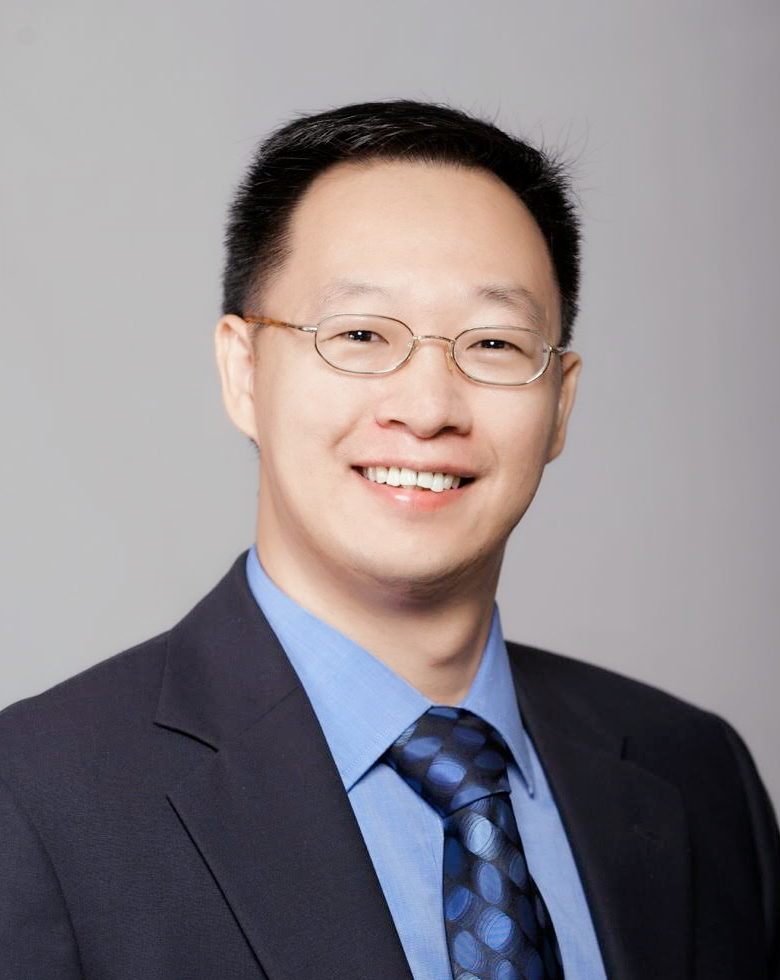}}]{Zizhong Chen}
(Senior Member, IEEE) received a bachelor's degree in mathematics from Beijing Normal University, a master's degree degree in economics from the Renmin University of China, and a Ph.D. degree in computer science from the University of Tennessee, Knoxville. He is a professor of computer science at the University of California, Riverside. 
His research interests include high-performance computing, parallel and distributed systems, big data analytics, cluster and cloud computing, algorithm-based fault tolerance, power and energy efficient computing, numerical algorithms and software, and large-scale computer simulations.
He currently serves as a subject area editor for \textit{Elsevier Parallel Computing} journal and an associate editor for the \textit{IEEE Transactions on Parallel and Distributed Systems}. Email: chen@cs.ucr.edu.
\end{IEEEbiography}

\vskip -2\baselineskip plus -1fil

\begin{IEEEbiography}[{\includegraphics[width=1in,height=1.25in,clip,keepaspectratio]{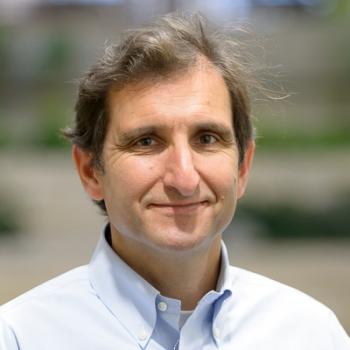}}]{Franck Cappello}
(Fellow, IEEE) is the director of the Joint-Laboratory on Extreme Scale Computing gathering six of the leading high-performance computing institutions in the world: Argonne National Laboratory, National Center for Scientific Applications, Inria, Barcelona Supercomputing Center, Julich Supercomputing Center, and Riken AICS. He is a senior computer scientist at Argonne National Laboratory and an adjunct associate professor in the Department of Computer Science at the University of Illinois at Urbana-Champaign. He is an expert in resilience and fault tolerance for scientific computing and data analytics. Recently he started investigating lossy compression for scientific data sets to respond to the pressing needs of scientist performing large-scale simulations and experiments. His contribution to this domain is one of the best lossy compressors for scientific data set respecting user-set error bounds. He is a member of the editorial board of the \textit{IEEE Transactions on Parallel and Distributed Computing} and of the \textit{ACM HPDC} and \textit{IEEE CCGRID} steering committees. He is a fellow of the IEEE. Email: cappello@mcs.anl.gov.
\end{IEEEbiography}




\end{document}

%% file: tex/abstract.tex
\begin{abstract}
Today's scientific simulations require a significant reduction of data volume because of extremely large amounts of data they produce and the limited I/O bandwidth and storage space. Error-bounded lossy compression has been considered one of the most effective solutions to the above problem. In practice, however, the best-fit compression method often needs to be customized or optimized in particular because of diverse characteristics in different datasets and various user requirements on the compression quality and performance.
In this paper, we address this issue with a novel modular, composable compression framework named SZ3. Our contributions are four-folds. (1) We develop SZ3 which features an innovative modular abstraction for the prediction-based compression framework, such that compression modules can be plugged in easily to create new compressors based on characteristics of data and user requirements. 
(2) We create a new compression pipeline by SZ3 for GAMESS data, which significantly improves the compression ratios over state-of-the-art compressors.
(3) We develop an adaptive compression pipeline by SZ3 for APS data with minimal efforts, which leads to the best rate-distortion among all existing error-bounded lossy compressors for any bit-rate. 
(4) We compare the sustainability of SZ3 with leading error-controlled prediction-based compressors, and then demonstrate the necessity of diverse pipelines by integrating and evaluating several compression pipelines on diverse scientific datasets from multiple disciplines. 
Experiments show that SZ3 incurs very limited overhead in compressor integration and our customized compression pipelines lead to up to 20\% improvement in compression ratios under the same data distortion, when compared with the best existing approach.
\end{abstract}

%% file: tex/introduction.tex
\IEEEraisesectionheading{\section{Introduction}\label{sec:introduction}}

\IEEEPARstart{D}{ata} reduction is becoming increasingly important to scientific research because of the large amount of data produced by simulations running on exascale computing systems and experiments conducted on advanced instruments. 
For instance, recent climate research, which performs climate simulation in 1 km$\times$1 km resolution, generates 260 TB of floating-point data every 16 seconds~\cite{foster2017computing}. 
When the generated data are dumped into parallel file systems or secondary storage systems to ensure long-term access, the limited storage capacity and/or I/O bandwidth will impose great challenges. 
While scientists aim to significantly reduce the size of their data to mitigate this problem, they are also concerned about the quality of data reduction.
General data reduction approaches, including traditional wavelet-based methods~\cite{jpeg, jpeg2000} and emerging neural-network-based methods~\cite{theis2017lossy, agustsson2019generative} widely used in the image processing community, may lead to loss of important scientific insights as they do not enforce quantifiable error bounds on reconstructed data. 

Over the past decade, error-bounded lossy compression~\cite{sz16, sz17, sz-reg, sz-autotuning, zfp, fpzip, ainsworth2017compression, ainsworth2018multilevel, ainsworth2019multilevelqoi, liang2021mgard+} has been proposed and employed to reduce scientific data while controlling the distortion. 
Depending on how the original data are decorrelated, existing compressors can be classified into prediction-based and transform-based. 
These compressors all allow users to specify an error bound during compression and ensure that the error between original and decompressed data is strictly than the bound. 
In this paper we focus mainly on prediction-based approaches because transformed-based approaches can be formulated to prediction-based ones by using the corresponding transforms as predictors (at the cost of certain speed degradation), as suggested by prior works~\cite{xin-sc19-hybrid-sz-zfp}.

Although existing prediction-based approaches such as SZ \cite{sz16,sz17, sz-reg} are general and can be applied to various scenarios, they may not lead to the best quality and performance given a specific dataset or error bound requirement. 
The best-fit compression method is never universal, which is true even for the same dataset because the compression efficiency would be affected by the required error bounds as well. 
For instance, SZ-1.4~\cite{sz17} with a Lorenzo predictor shows very good compression ratios with low error bounds, but it suffers from low quality and artifacts with high error bounds, where approaches with a regression-based predictor~\cite{sz-reg} or an interpolation-based predictor~\cite{zhao2021optimizing} have been proved to be much more efficient.
Likewise, data generated by the GAMESS quantum chemistry package~\cite{alexeev2012gamess} exhibits periodic scaled patterns, where a pattern-based predictor demonstrates obvious improvements in both compression speed and ratios~\cite{pastri}. 
Thus, a loosely coupled compression framework that allows for customization of the prediction-based error-bounded lossy compression model is critical to optimizing the compression quality and performance for users in practice. 

In this paper, we present a modular and composable framework---SZ3---which can be used to easily create new error-bounded lossy compressors on demand. 
SZ3 features a modular abstraction for the prediction-based compression pipelines such that modules can be developed and adopted independently. Specifically, users can customize any stages in the compression pipeline, including preprocessing, prediction, quantization, encoding, and lossless compression, via carefully designed modules. 
Based on these customized modules, SZ3 allows users to compose their own compressors (or compression pipelines) to adapt to diverse data characteristics and requirements, thus achieving high compression quality and performance with minimal effort. Such a composable design is able to provide a variety of useful supports, including point-wise relative error bounds (logarithmic transform-based preprocessor~\cite{xincluster18}), feature-preserving compression (element-wise quantizer~\cite{liang2020toward}), and speed-ratio tradeoffs (module bypass). 
Although designed for data in Cartesian grids, SZ3 can also work with data in unstructured grids by applying a linearization which re-arranges data to a one-dimensional array. 

We summarize our contributions as follows.
\begin{itemize}
\item We carefully design and develop SZ3, a flexible, efficient framework that allows easy creation and customization of prediction-based error-bounded lossy compressors. This work is critical to obtaining high data compression quality because of diverse scientific data characteristics and user requirmeents in practice. 
\item We develop a new compressor using SZ3 for data generated from GAMESS quantum chemistry package. By substituting the default quantizer with a specialized one and augmenting a lossless compression stage, the composed compressor achieves better performance than current state of the art with minimal effort.
\item We develop an efficient compressor using SZ3 for data collected from Advanced Photon Source instruments. By incorporating an adaptive pipeline with existing modules, the composed compressor leads to the best rate-distortion under any bit rate.
\item We compare the sustainability of SZ3 with leading prediction-based compressors, and then integrate several compression pipelines to demonstrate the necessity of diverse pipelines. The performance and efficiency are carefully characterized using diverse scientific datasets across multiple domains.
\end{itemize}

The rest of the paper is organized as follows. In Section~\ref{sec:related} we discuss  related work. In Section~\ref{sec:design} we present the design and modules of SZ3 framework. In Section~\ref{sec:gamess} and Section~\ref{sec:aps} we describe how we leverage the proposed framework to create efficient compressors for GAMESS and APS data in details.
In Section~\ref{sec:pipelines} we present the comparison on sustainability and evaluation for diverse pipelines.
In Section~\ref{sec:conclusion} we conclude with a vision of future work.     

%% file: tex/related.tex
\section{Related Work}
\label{sec:related}
With more powerful high-performance computing (HPC) systems and high-resolution instruments, the volume and generation speed of scientific data have been experiencing an unprecedented increase in recent years, causing problems in data storage, transmission, and analysis. 
Compared with the fast evolution of computing resources, the I/O systems are heavily underdeveloped, remaining  a bottleneck in most scenarios. 
Data compression is regarded as a  direct way to mitigate such a bottleneck, and many approaches have been presented in the literature to address this issue.

Lossless compressors~\cite{gzip, zstd, fpc, spdp, alted2017blosc} ensure that no information is lost during the compression. 
Despite their success in many fields, lossless compressors suffer from low compression ratios on floating-point scientific data due to the almost randomly distributed mantissas. 
Previous work~\cite{lindstromerror} has shown that state-of-the-art lossless compressors can lead to a compression ratio of only 2 when directly applied to most floating-point scientific datasets, whereas scientific applications usually require over $10\times$ reduction on their data~\cite{use-case}. 

Lossy compressors~\cite{jpeg,jpeg2000, theis2017lossy, agustsson2019generative, li2019vapor, ma2020end} offer the flexibility to trade off data quality for high compression ratios, but they may result in a higher distortion than users' expectation. 
The unbounded distortion may result in unexpected behaviors in post hoc data analytics and even false discoveries, leaving risks in trusting the analysis results on the decompressed data.

In comparison with traditional lossy compression, error-bounded lossy compression has been rapidly developed to fill the gap by reducing the size of scientific data while guaranteeing quantifiable error bounds. 
Prediction-based and transform-based models are the most popular models for designing error-bounded lossy compressors. 
One of the most well-known transform-based error-bounded lossy compressors is ZFP~\cite{zfp}, which decorrelates the data using a near-orthogonal transform and encodes the transformed coefficients using embedded encoding. 
MGARD~\cite{ainsworth2017compression, ainsworth2018multilevel, ainsworth2019multilevelqoi} is another compressor relying on the transform-based model. It leverages wavelet theories and $L^2$ projection for data decorrelation, followed by linear-scaling quantization, variable-length encoding, and lossless compression. 

According to recent studies~\cite{understand-compression-ipdps18}, SZ~\cite{sz16, sz17, sz-reg} is regarded as one of the leading prediction-based lossy compressor in the scientific computing community. SZ follows a 4-step pipeline to perform the compression, namely data prediction, quantization, Huffman encoding, and lossless compression. 
Significant efforts have been made to enable new features or functionalities based on this pipeline. 
For instance, in~\cite{xincluster18}, a logarithmic transform was used in a preprocessing step to change a pointwise-relative-error-bound compression problem to an absolute-error-bound compression problem, which is then solved by the SZ compression pipeline. 
In~\cite{liang2020toward}, the authors derived the element-wise error bounds based on how critical points are extracted, and they leveraged the SZ compression pipeline along with element-wise quantization to ensure that those critical points are preserved in the decompressed data. 
In~\cite{pastri}, the authors adjusted the pipeline by using a pattern-based predictor to better exploit the correlation in data and a predefined fixed Huffman tree for faster encoding. 
Attempts were also made to use the near-orthogonal transform in ZFP as a predictor in the pipeline~\cite{xin-sc19-hybrid-sz-zfp}. 
All the above works, however, are developed within a tightly-coupled design, so that the compression pipelines cannot be adjusted on demand, which thus cannot adapt to user's diverse requirements or different use-cases in turn. By contrast, the SZ3 framework offers a breakthrough, flexible, modular framework, which can be leveraged to adapt to diverse use-cases very efficiently.

Although many efforts have been spent on abstracting lossy compression, most of them are focused on enabling an adaptive selection of existing compressors.
For instance, SCIL~\cite{kunkelDecouplingSelectionCompression2017} attempts to abstract across compressors and acts as a metacompressor that provides backends to various existing algorithms.
LibPressio~\cite{fraz} provides a common API for different compressors to allow for easy integration of lossy compression in an extensible fashion. 
Instead, SZ3 separates and abstracts stages in the prediction-based compression model, allowing for easy creation of new compressors in fine granularity rather than selection of existing ones. 
To the best of our knowledge, this is the first attempt to build a generic framework that allows users to easily customize their own compressors based on their actual needs.

%% file: tex/design.tex
\section{SZ3: A Modular Compression Framework}
\label{sec:design}
\begin{figure*}
    \centering
    \includegraphics[width=\textwidth]{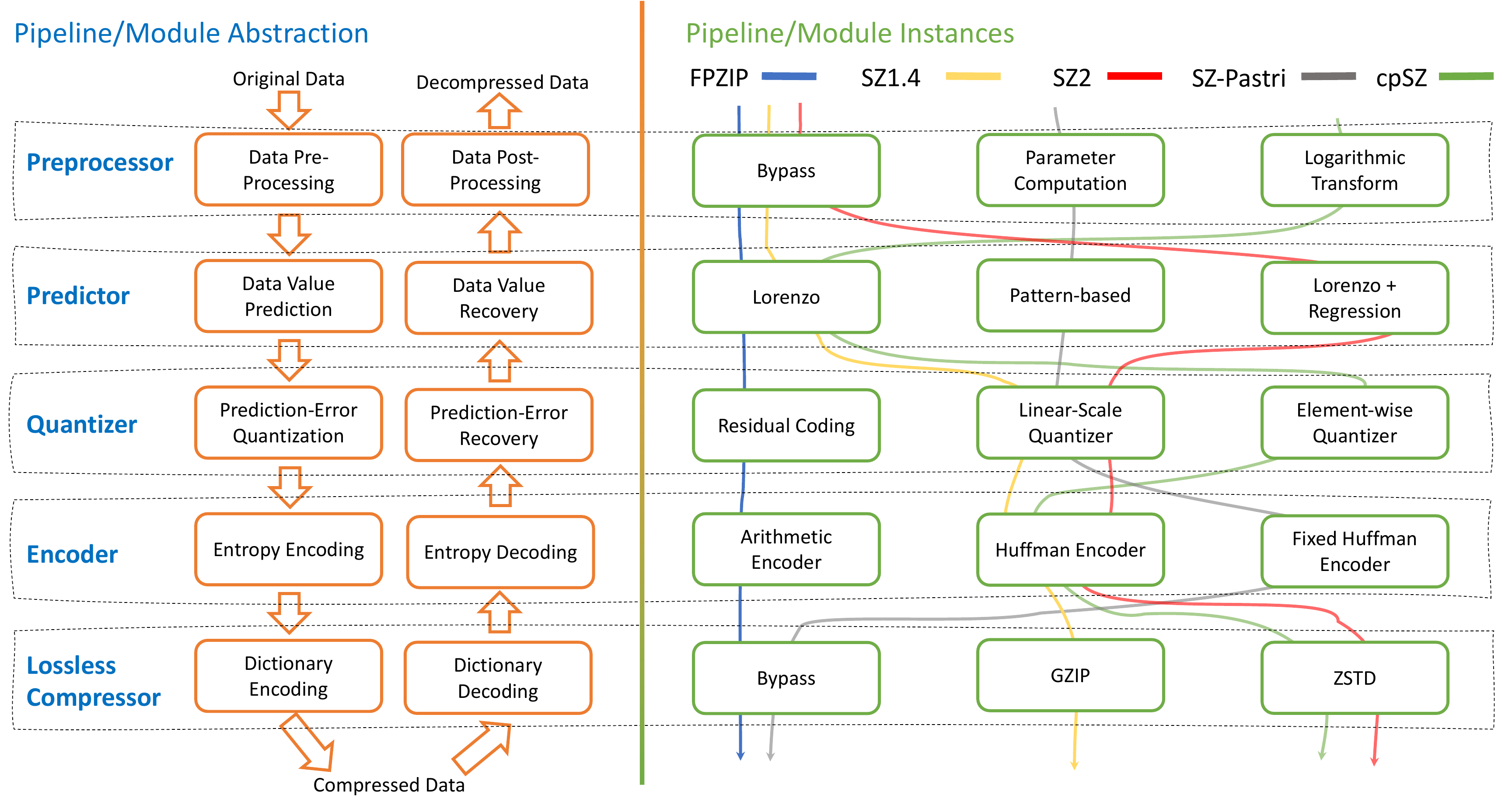}
    \vspace{-1.5em}
    \caption{SZ3 design overview: left part of the figure shows the abstraction and key functionalities of prediction-based compression pipeline with SZ3 modules; right part of the figure displays common instances of these modules and how five leading compressors are composed by these instances.}
    \label{fig:sz3-arch}
    \vspace{-1em}
\end{figure*}

In this section we introduce the design and implementation of SZ3.
With modularity in mind, SZ3 enables easy customization of prediction-based compression pipelines with minimal overhead.

\subsection{Design overview}
Figure~\ref{fig:sz3-arch} illustrates the design overview of SZ3. 
The compression process is abstracted into five stages (displayed as the dotted boxes), each of which serves as an individual module. 
Orange boxes depict the key functionalities of each module and green boxes illustrate several corresponding instances. 
A compressor is realized by identifying a compression pipeline which is composed by instances from each module. 
This figure demonstrates how five leading compressors designed for different purposes, namely FPZIP~\cite{fpzip}, SZ1.4~\cite{sz17}, SZ2~\cite{sz-reg}, SZ-Pastri~\cite{pastri}, and cpSZ~\cite{liang2020toward}, are composed using this abstraction (see the solid lines), which shows the generality of the abstraction. 
For instance, the FPZIP compression pipeline bypasses the precessor and leverages Lorenzo predictor for data decorrelation, followed by residual encoding to ensure error control and arithmetic encoding for size reduction. 
In the following text we will detail the modular design in SZ3, along with example instances of the modules.

\subsection{Modularity}
In this section we discuss the five modules in SZ3, namely preprocessor, predictor, quantizer, encoder, and lossless compressor, with module instances that have proven to be effective for scientific datasets. 
Developers can write their own module instances and plug them in the compression pipeline to design prediction-based error-bounded lossy compression for their dataset. 
Due to space limitation, we present only the most important functions and several representative instances for each module. Detailed interfaces for each module are listed in Appendix A.


\textbf{Preprocessor (see Appendix \ref{appd:preprocess}):}~
The preprocessor is used to process the input dataset for achieving high efficiency or diverse requirements before performing the actual compression. 
The key function in the preprocessor, namely \texttt{preprocess}, takes in original data and compression configuration as input, and then transforms the data in an in-place fashion and change the compression configuration accordingly. 
If users want to keep original data while the preprocessor needs to alter the data, a separate buffer is required to perform the preprocessing. 
Based on the actual design, the \texttt{postprocess} function either reverses the preprocessing procedure or is omitted. 

\textit{Instances:} A typical preprocessor for error-controlled lossy compressors is the logarithmic transform used to enable point-wise relative error bounds~\cite{xincluster18}, where data are transformed to the logarithmic domain and compressed with an absolute error bound transformed from the point-wise relative one. 
Besides, SZ-Pastri~\cite{pastri} requires a preprocessing step to identify the proper parameters, such as block size and pattern size, for the pattern-based predictor.
In Section~\ref{sec:aps}, we further leverage a preprocessor to alter the layout of data for better compression ratio. This is based on our observations that some 3D datasets will have a better compression ratio when treated as a 2D or 1D dataset (as will be detailed later). 



\textbf{Predictor (see Appendix \ref{appd:predictor}):}~
Predictors are the key components of prediction-based compressors, which perform value prediction based on diverse patterns for data decorrelation.
There are two important functions in the predictor interface, namely \texttt{predict} and \texttt{save}/\texttt{load}. The \texttt{predict} function outputs the predicted value based on the characteristics of the underlying predictor using the multidimensional iterator (to be detailed in Section~\ref{sec:sustainability}).  
Necessary information about the predictor, for instance the coefficients of the regression predictor~\cite{sz-reg,sz-autotuning}, will be recorded in the \texttt{save} function. 
During decompression, \texttt{load} function will be invoked to reconstruct the predictor.

\textit{Instances: } Lorenzo predictor~\cite{lorenzo} and its high order variations~\cite{sz17}, which perform multidimensional prediction for each data point based on its neighbor data points, are classic and powerful prediction methods used in lossy compressors such as SZ~\cite{sz17} and FPZIP~\cite{fpzip}.
In~\cite{sz-reg}, a regression-based predictor is proposed to construct a hyperplane and uses points on the hyperplane as predicted values, which significantly improves the prediction efficiency when user-specified error bound is high. 
We further implement a composite predictor instance inherited from this interface, which may consist of multiple predictors using different prediction algorithms. 
This requires an error estimation function for each predictor, which will be used to determine the best-fit predictor for a given data chunk. 
The statistical approach in~\cite{sz-reg} and~\cite{liang2021mgard+} is generalized as the estimation criterion in SZ3. 
With the composite predictor, multialgorithm designs with more than one predictors can be implemented very easily.

\textbf{Quantizer (see Appendix \ref{appd:quantizer}):} The quantizer is used to approximate prediction errors generated by the predictors with a smaller countable set to reduce their entropy while respecting the error bound. 
As the only module that introduces errors in the compression pipeline, quantizer determines how the final errors in the decompressed data are controlled. 
The \texttt{quantize} function is the most important function in a quantizer, where the prediction error is quantized based on the original data value and its predicted value from the predictor. 
During decompression, the decompressed data value is computed by the \texttt{recover} function, which reverses the steps in the \texttt{quantize} function. 
The quantizer module is also responsible for encoding/decoding the unpredictable data, i.e., data fall out of the countable set. This is realized in the \texttt{save}/\texttt{load} function.

\textit{Instances: } Linear-scaling quantizer~\cite{sz17} is a widely used quantizer to enable absolute error control in lossy compression. 
In particular, this quantizer constructs a set of equal-sized consecutive bins each with twice the error bound in length. 
Then, the prediction error will be translated into the index of the bin containing it.
Prediction errors that fall out of range are regarded as unpredictable and will be encoded and stored separately. 
Besides, log-scale quantizer~\cite{numarck} is used to adjust the size of bins for a more centralized error distribution and element-wise quantizer~\cite{liang2020toward} is used to provide fine-granularity error control for each data point.

\textbf{Encoder (see Appendix \ref{appd:encoder}):} Encoder is a lossless scheme to reduce the storage of integer indices (or symbols) generated by quantizers. 
The encoder module involves two essential functions---\texttt{encode} and \texttt{save}/\texttt{load}. 
The \texttt{encode} function transforms the quantized integers from the quantizer to compressed binary formats; similar to other modules, the encoder module has a \texttt{decode} function which performs the reverse process during decompression. 
This module also has \texttt{save}/\texttt{load} functions for storing/recovering metadata such as the Huffman tree. 

\textit{Instances:} Huffman encoder~\cite{huffman} is a classic variable-length encoding algorithm that uses fewer bits to represent more common symbols. This encoder first constructs a Huffman tree based on the frequency of input data using a greedy algorithm, generates codebook according to the tree, and then compress the data using the codebook. 
The fixed Huffman encoder used in SZ-Pastri~\cite{pastri} is a variation of the Huffman encoder, which uses a predefined Huffman tree instead of constructing one on the fly to eliminate the cost for both construction and storage of the tree.
Arithmetic encoder is another type of encoder widely used in data compression, which represents current information as range and encodes the entire data into a single number.

\textbf{Lossless Compressor (see Appendix \ref{appd:lossless}):} Lossless compressors
are used to further shrink the size of compressed binary formats produced by the encoders, because the entropy-based encoders may overlook repeated patterns in the data thus lead to suboptimal compression ratios. 
The lossless compressor module in SZ3 acts mainly as a proxy of state-of-the-art lossless compression libraries.
This module invokes external libraries to compress the output from the encoder module with \texttt{compress} and \texttt{decompress} interfaces. 

\textit{Instances: } We provide portable interfaces in SZ3 to integrate with state-of-the-art lossless compressors including ZSTD~\cite{zstd}, GZIP~\cite{gzip}, and BLOSC~\cite{alted2017blosc}. Because lossless compressor is a standlone module attached to the previous stages, it would be fairly easy to include and integrate new lossless compression routines as well. 

\subsection{Compression pipeline composition}
In SZ3, a compression pipeline can be composed by identifying the instances of modules and putting them together. Algorithm~\ref{alg:compression} shows how a general-purpose error-controlled lossy compressor is composed using the selected preprocessor, predictor, quantizer, encoder, and lossless compressor. 
In addition, SZ3 employs compile time polymorphism (see Section~\ref{sec:sustainability}) such that users can switch the instances without bothering to modify the compression functions. 
This makes SZ3 highly adaptive to diverse use cases, with significantly reduced efforts on compressor development. 

\begin{algorithm}[ht]
\caption{\textsc{A General Compressor in SZ3}} \label{alg:compression} \footnotesize
\renewcommand{\algorithmiccomment}[1]{/*#1*/}
\textbf{Input}: input data $d$ of size $n$, compression configuration $conf$\\
\textbf{Output}: compressed data $cc$

\begin{algorithmic} [1]
\STATE $preprocessor.\texttt{process}(d, conf)$ \COMMENT{perform preprocessing}
\FOR{$i = 1 \to n$}
	\STATE $p \gets predictor.\texttt{predict}(d[i])$   \COMMENT{perform prediction}
	\STATE $q[i] \gets quantizer.\texttt{quantize}(d[i], p)$    \COMMENT{perform quantization}
\ENDFOR
\STATE $c \gets \texttt{allocate\_memory}()$ 
\STATE $predictor.\texttt{save}(c)$ \COMMENT{save predictor}
\STATE $quantizer.\texttt{save}(c)$ \COMMENT{save quantizer}
\STATE $encoder.\texttt{encode}(q, c)$ \COMMENT{perform encoding}
\STATE $encoder.\texttt{save}(c)$ \COMMENT{save encoder}
\STATE $cc \gets lossless\_compressor.\texttt{compress}(c)$ \COMMENT{perform lossless compression}
\RETURN $cc$
\end{algorithmic}
\end{algorithm}

%% file: tex/usecases.tex
\section{Developing an Efficient Compressor for GAMESS Data using SZ3}\label{sec:gamess}
In this section, we present how we create a new compressor using SZ3, which can improve the compression ratios for the data generated from the real-world scientific simulation GAMESS~\cite{alexeev2012gamess}. In the following text, we first introduce the GAMESS data and its current compressor --- SZ-Pastri~\cite{pastri}, and then present our characterization on the quanzation integers and the new customization method. At last, we evaluate the compression ratios and speed based on three representative data fields in GAMESS.



\input{tex/usecases/pastri}

\section{Composing an Efficient Compressor for APS Data using SZ3}\label{sec:aps}
\input{tex/usecases/aps}

\input{tex/usecases/pipeline}

%% file: tex/usecases/pastri.tex
\subsection{GAMESS data and SZ-Pastri Compressor}
\label{sec:pastri}

Quantum chemistry researchers often need to obtain a wavefunction by solving the Schr\"odinger differential equation, which involves all the chemical system's  information. The wavefunction needs to be constructed by two-electron repulsion integrals (ERI), which requires too large a memory capacity to hold at runtime during the simulation. A straightforward solution is reproducing the ERI dataset whenever needed during the simulation, although this would significantly delay the simulation because of the fairly expensive cost in generating the ERI data. In our prior work, we developed an efficient error-bounded lossy compressor called SZ-Pastri~\cite{pastri}, which can compress the ERI data in memory and decompress it in the beginning of each iteration of the simulation. Such a method can effectively avoid the ERI recalculation cost, so as to improve the overall performance. 
SZ-Pastri takes advantages of the periodic patterns that exist in the GAMESS dataset, because the ERI values are calculated in order and are dependent on shape and distance of electron clouds. 
Specifically, SZ-Pastri identifies a periodic pattern and uses it along with a scaling coefficient for each block to enable accurate data prediction.
This leads to substantial performance gain compared to existing general compressors~\cite{sz-reg, zfp}.

\subsection{Data characterization and pipeline customization}
\begin{figure*}[ht]
\centering
\subfigure[Data]{
\raisebox{-1cm}{\includegraphics[width=0.6\columnwidth]{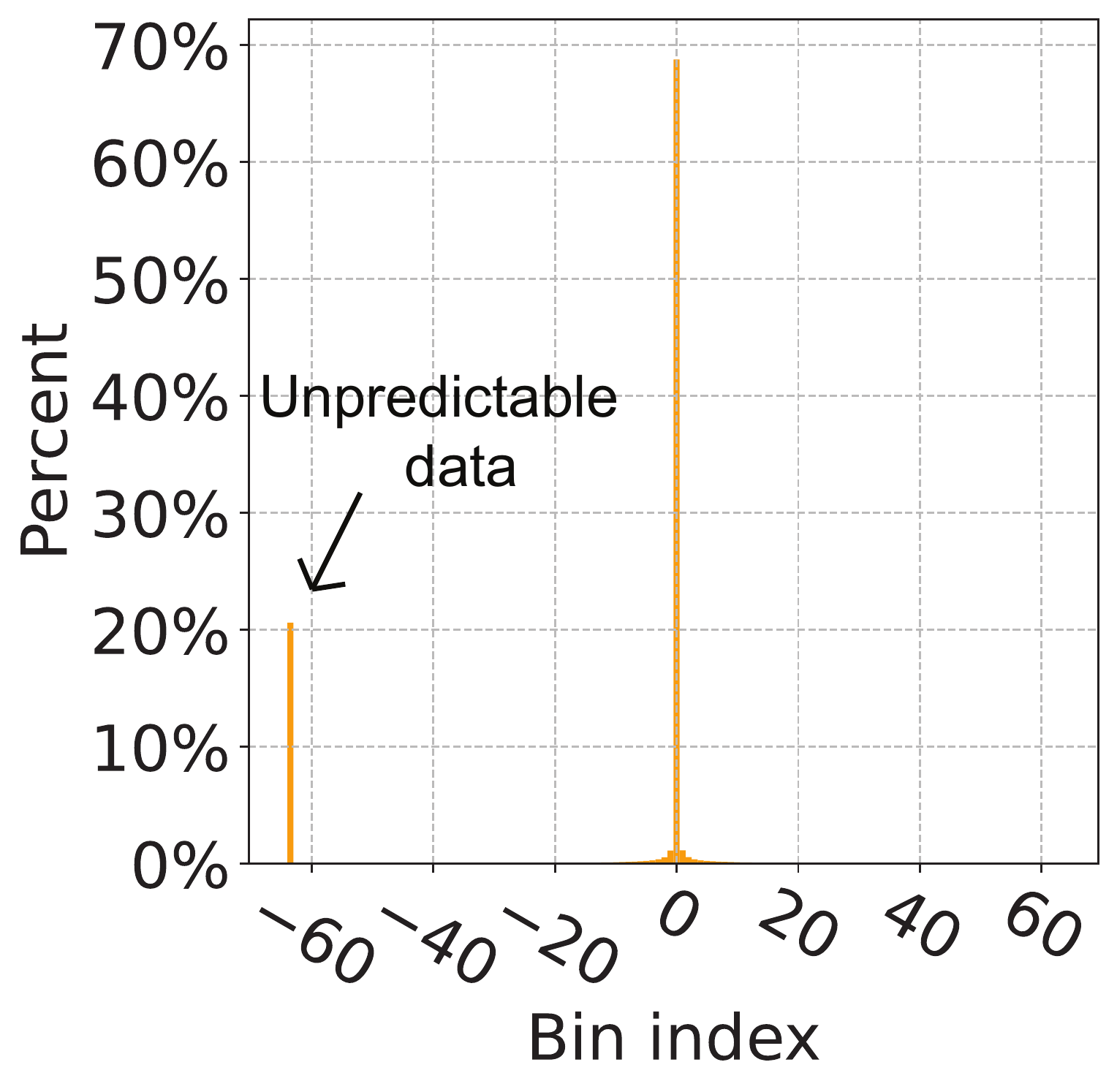}}}
\hspace{-2mm}
\subfigure[Pattern]{
\raisebox{-1cm}{\includegraphics[width=0.6\columnwidth]{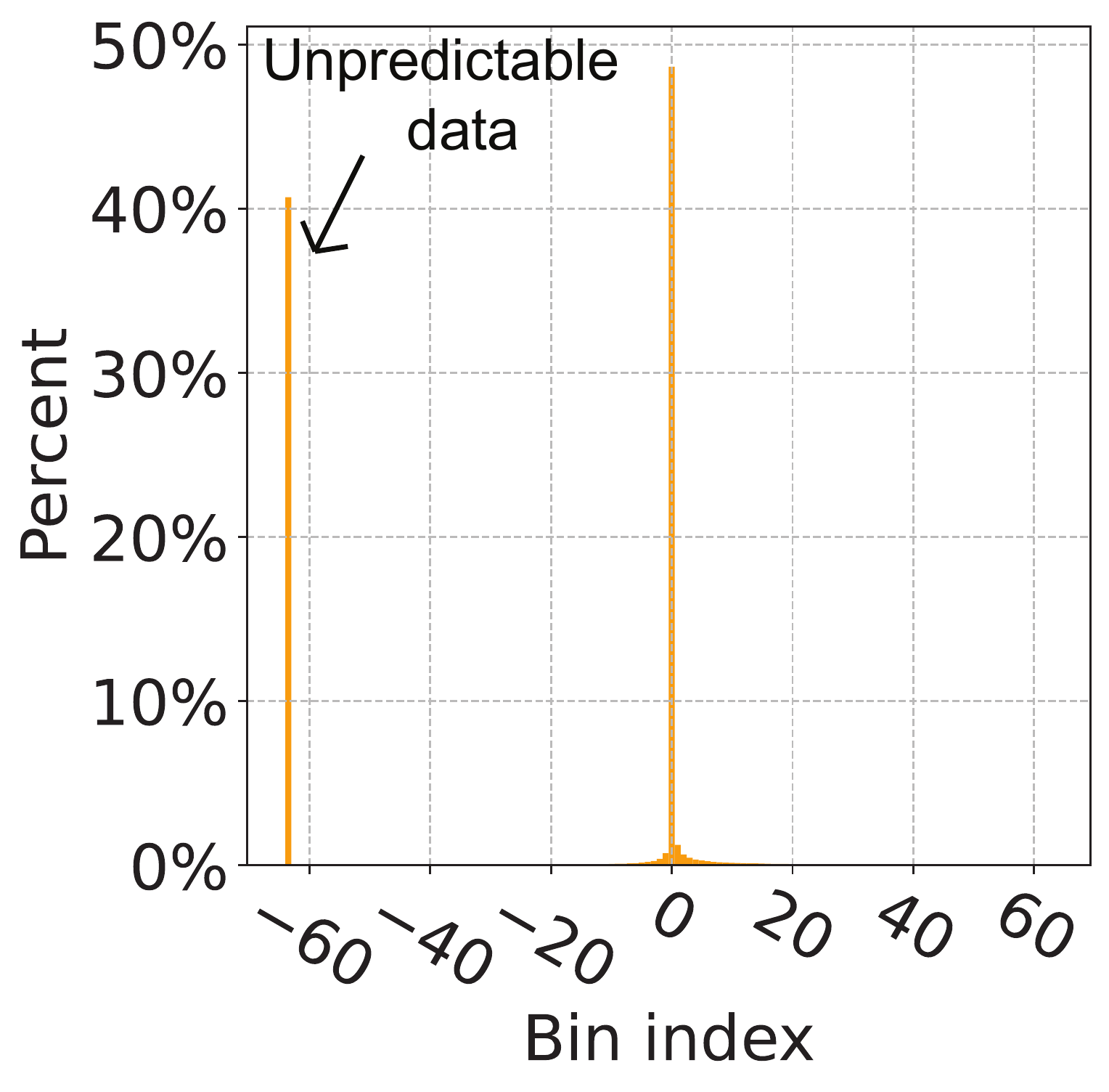}}}
\hspace{-2mm}
\subfigure[Scale]{
\raisebox{-1cm}{\includegraphics[width=0.6\columnwidth]{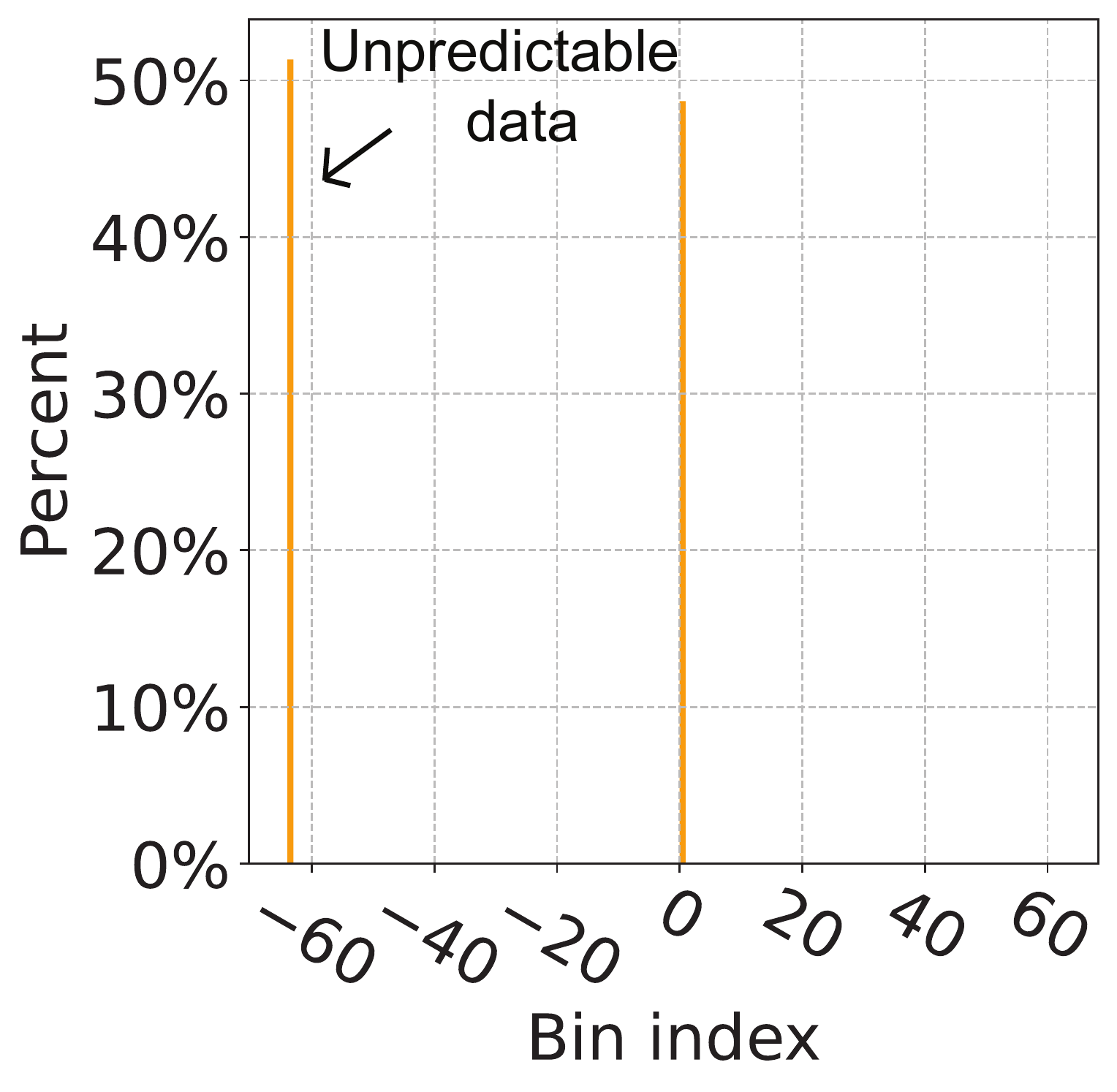}}}
\vspace{-1em}
\caption{Distribution of quantization integers in SZ3-Pastri.}
\label{fig:pastri-distribution}
\vspace{-1em}
\end{figure*}

We first characterize the quantization integers for SZ-Pastri, which are the most impactful factors for the final compression ratios. 
To enable correct decompression, SZ-Pastri needs to quantize and store the information for both the periodic patterns and block-wise scales. 
Thus, the quantization integers in SZ-Pastri consist of three components, which are computed from data, patterns, and scales, respectively.
As displayed in Figure~\ref{fig:pastri-distribution}(a), the distribution of quantization integers for the pattern-based predictor is centered in $0$, which indicates very high prediction accuracy and thus better compression ratios. 
However, a significant percentage (20\% for data) of the quantization integers fall out of the quantization range ($64$ in this setting).
These data, usually described as unpredictable, require additional mechanisms for storage in order to be correctly recovered during decompression.  
In SZ-Pastri, they are directly truncated and stored based on the user-specified error, which fails to exploit the correlation in the data to achieve high compression, although relatively fast compression speed is provided. 


\begin{figure}[ht]
\centering
\includegraphics[width=0.55\columnwidth]{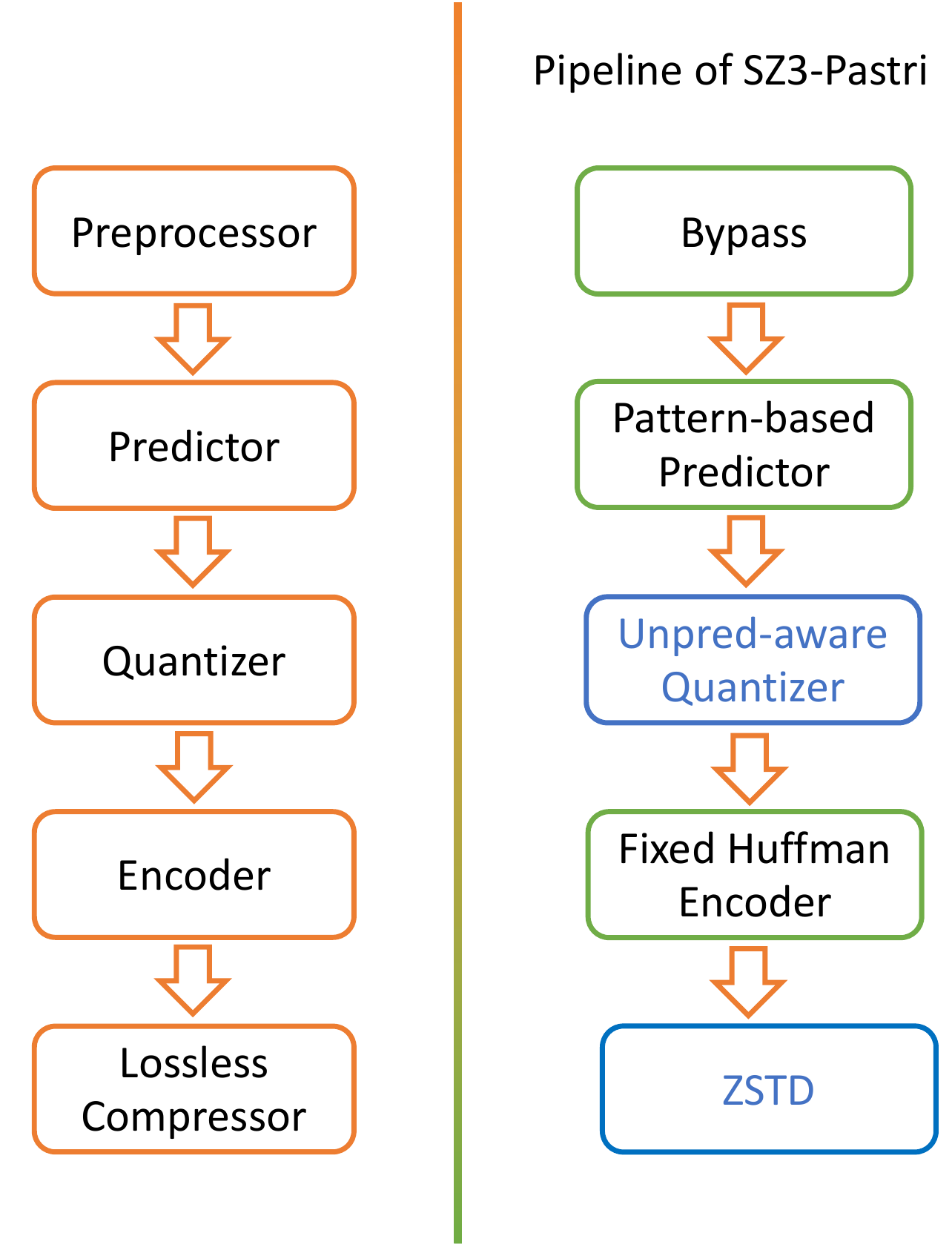}
\vspace{-1em}
\caption{Compression pipelines for GAMESS data. Blue boxes indicate optimized/added modules in SZ3-Pastri over SZ-Pastri.}
\label{fig:pastri-pipeline}
\end{figure}

Based on these observations, we improve the compression efficiency of SZ-Pastri by leveraging a specialized quantizer to deal with the unpredictable data. 
Inspired by the embedded encoding approaches widely used in transform-based compressors~\cite{zfp, liang2021error}, we store data in the order of bitplane instead of applying the truncation directly. 
A bitplane represents a set of bits corresponding to a given bit position in the binary representations of the data. 
Because small data values have meaningful bits only in less significant bitplanes, the relatively significant bitplanes will yield good compression ratios because of consecutive $0$s. 
Similar to~\cite{zfp}, we first align the exponents of the prediction difference on unpredictable data to that of the error bound to convert the floating-point data into integers. 
These integers are then recorded in the order of bitplanes, namely, from the most significant bitplane to the least significant bitplane. 
Compared with direct truncation, this encoding method will not change the encoded size at this stage; however, its compressive encoded format will promise better compression ratios when lossless compression is adopted.  
Since this quantizer takes special care of unpredictable data storage, we name it \textit{Unpred-aware Quantizer} throughput the paper.
To take advantage of this method, we also add a lossless stage to the composed compression pipeline, as displayed in Figure~\ref{fig:pastri-pipeline}. This new compressor is called SZ3-Pastri, as it optimizes SZ-Pastri using the SZ3 framework. 

\subsection{Evaluation results}
We evaluate our method and compare it with SZ-Pastri and its variation (SZ-Pastri equipped with lossless compression) using three representative fields in GAMESS. 
Unless otherwise noted, all the experiments in this paper are conducted on the Bebop supercomputer~\cite{bebop} at Argonne National Laboratory. Bebop has 664 Broadwell nodes, each of which is equipped with two Intel Xeon E5-2695v4 processors containing 36 physical cores in total and 128 GB of DDL4 memory.

\begin{figure*}[ht]
\vspace{-0.5em}
\centering
\subfigure[{$ff|ff$}]
{
\raisebox{-1cm}{\includegraphics[width=0.55\columnwidth]{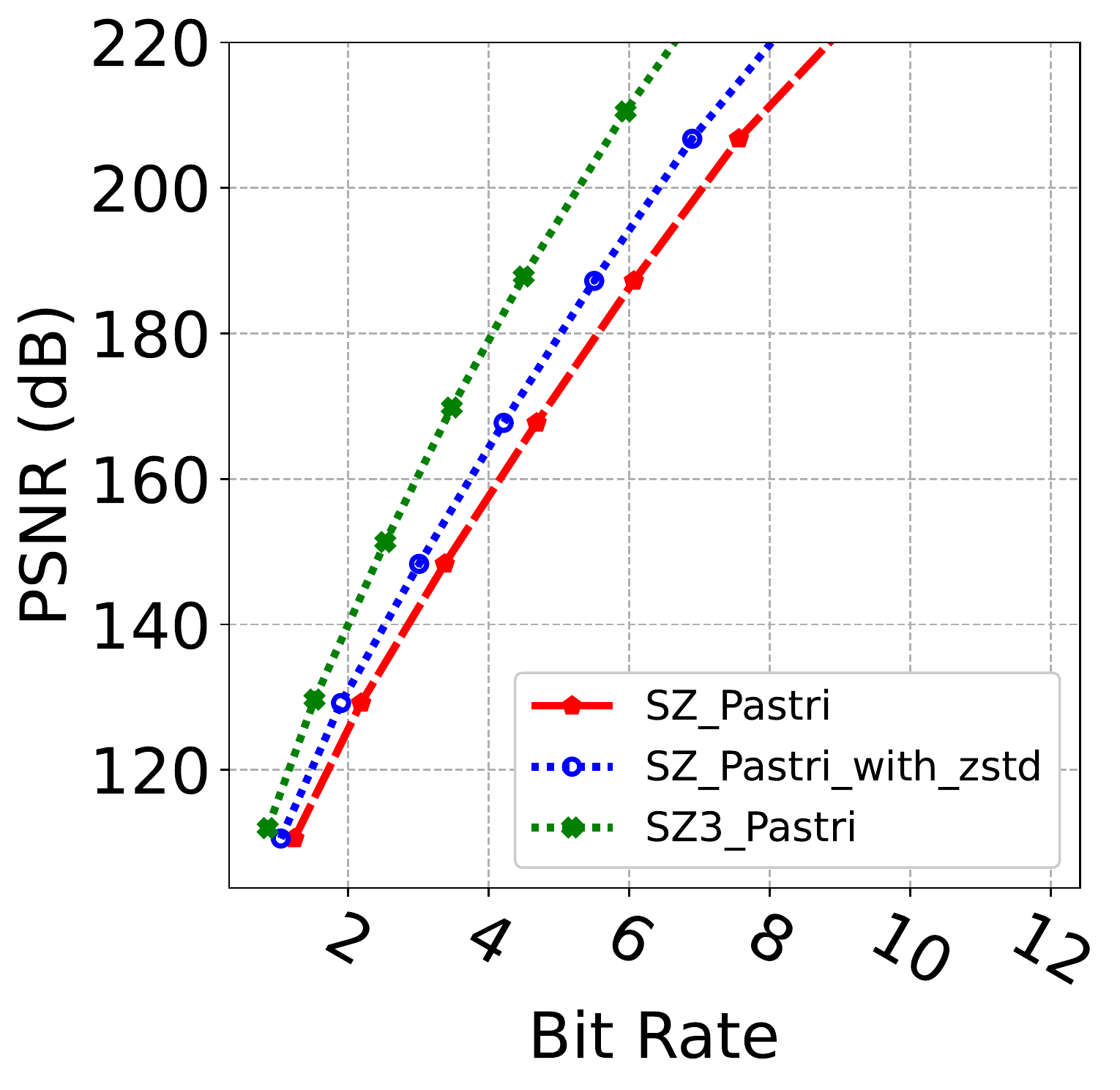}}
}
\subfigure[{$ff|dd$}]
{
\raisebox{-1cm}{\includegraphics[width=0.55\columnwidth]{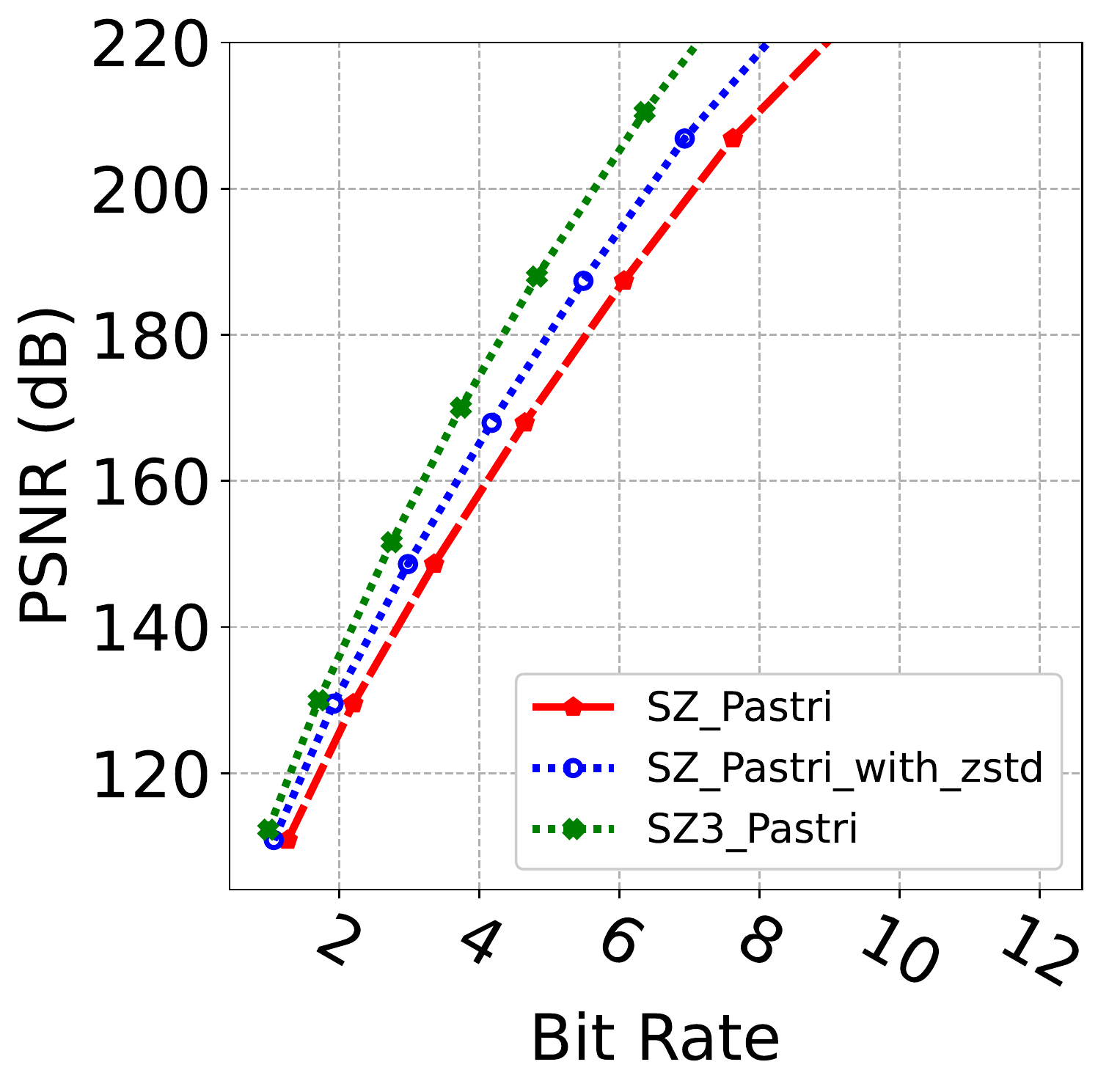}}
}
\subfigure[{$dd|dd$}]
{
\raisebox{-1cm}{\includegraphics[width=0.55\columnwidth]{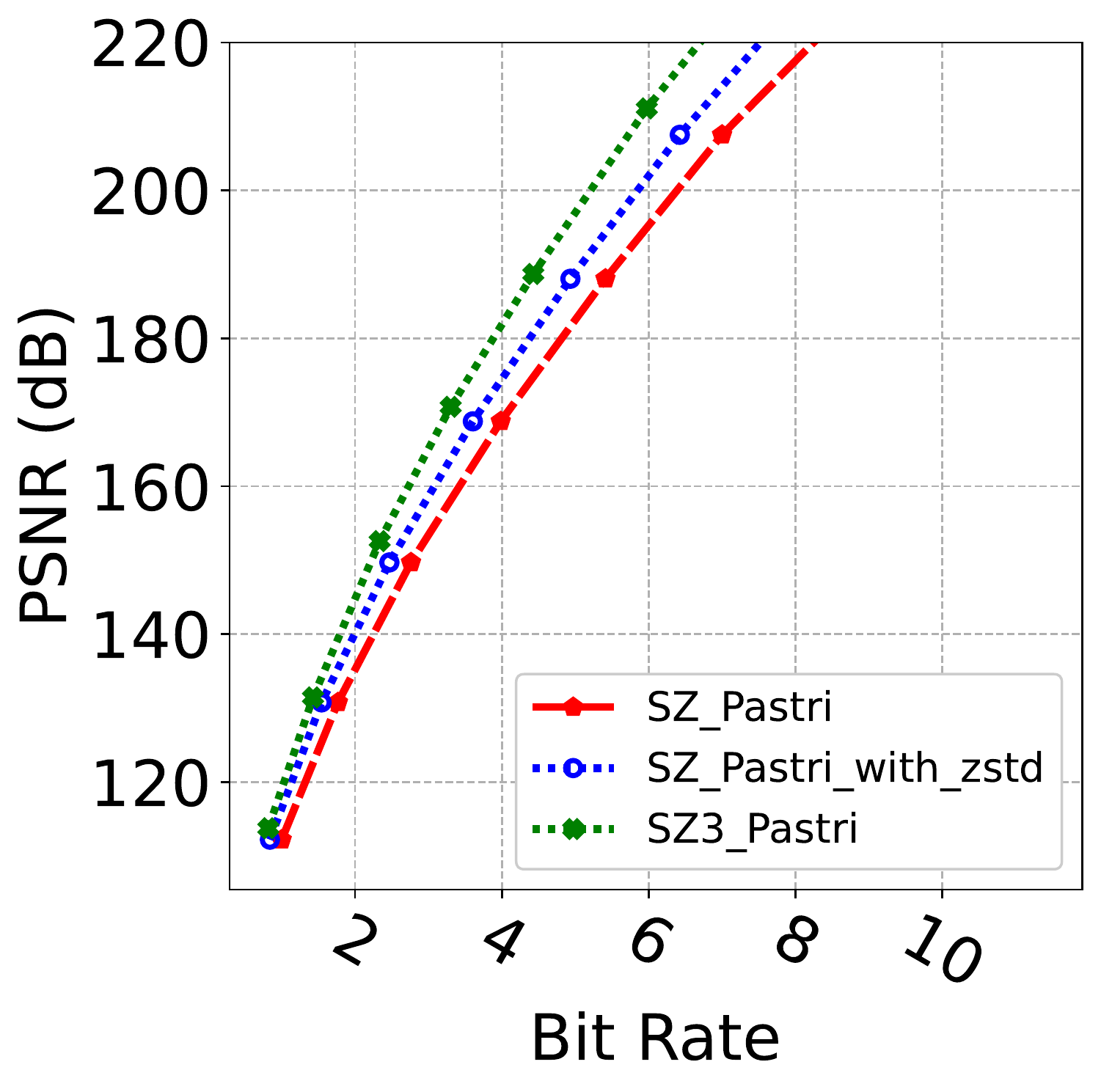}}
}
\vspace{-1em}
\caption{Rate-distortion on GAMESS data.}
\label{fig:pastri-rate-distortion}
\vspace{-1em}
\end{figure*}

The rate-distortion graphs of the evaluation are displayed in Figure~\ref{fig:pastri-rate-distortion}.
This graph entails the correlation between bit rate and Peak Signal-to-Noise Ratio (PSNR). The bit rate equals $bits/cr$ where $bits$ is number of bit in original data representation (e.g., $32$ for single-precision and $64$ for double-precision floating-point data) and $cr$ is the compression ratio. PSNR is inversely proportional to the mean square error of decompress data and original data in logarithmic scale.
Lower bit rate and higher PSNR indicate better compression quality.
According to this figure, SZ3-Pastri leads to the best rate-distortion along almost all bit rates. 
For example, the improvements of compression ratios on the $ff|ff$ dataset are generally $40\%$ and $20\%$, respectively, compared with SZ-Pastri and its lossless variation. 
We also show the exact compression ratio and speed of the three approaches under the desired absolute error tolerance (1E-10 according to the domain scientists) in Table~\ref{tab:result-GAMESS}. 
Compared with original SZ-Pastri, SZ3-Pastri significantly improves the compression ratios under the requirements. 
However, it has a degradation in performance, which is caused by the embedded encoding on unpredictable data (i.e., unpred-aware Quantizer, which improves the compression ratio) and the final lossless compression. 

\begin{table}[ht]\centering
\vspace{-1em}
\caption{Result on GAMESS data when absolute error bound is 1E-10}
\label{tab:result-GAMESS}
\vspace{-2mm}
\resizebox{\linewidth}{!}{
\begin{tabular}{|c|c|c|c|c|}
\hline
Dataset & Compressor & Ratios & Compression Speed\\
\hline
\multirow{3}{*}{$ff|ff$}  & SZ-Pastri & 8.46 & 662.01 MB/s \\
\cline{2-4}
 & SZ-Pastri-with-zstd & 9.27 & 377.17 MB/s \\
\cline{2-4}
 & SZ3-Pastri & 10.76 & 244.43 MB/s\\
\hline
\multirow{3}{*}{$ff|dd$}  & SZ-Pastri & 8.40 & 643.58 MB/s \\
\cline{2-4}
 & SZ-Pastri-with-zstd & 9.23 & 370.88 MB/s \\
\cline{2-4}
 & SZ3-Pastri & 10.06 & 221.03 MB/s \\
\hline
\multirow{3}{*}{$dd|dd$}  & SZ-Pastri & 9.14 & 613.12 MB/s\\
\cline{2-4}
 & SZ-Pastri-with-zstd & 9.96 & 364.51 MB/s \\
\cline{2-4}
 & SZ3-Pastri & 10.71 & 226.80 MB/s\\
\hline
\end{tabular}
}
\vspace{-1em}
\end{table}

%% file: tex/usecases/aps.tex
We then leverage our SZ3 framework to create an adaptive compression pipeline for the X-ray ptychographic data acquired at the Advanced Photon Source (APS). Similar to the previous section, we first introduce APS data, followed by the data characterization and compression pipeline customization along with the evaluation.

\subsection{APS data}
X-ray ptychography is a main high-resolution imaging technique that takes advantage of the coherence provided by the synchrotron source. However, this computational method of microscopic imaging requires much larger data volume and computational resource compared with conventional microscopic techniques. A revolutionary increase of about 3 orders of magnitude in the coherent flux provided by the coming APS upgrade will aggravate the burden of the data transfer and storage. Therefore, a new compression strategy with high compression ratios is being highly pursued in ptychography. In order to represent most sample scenarios, two ptychographic datasets were acquired from a computer chip pillar (isolated sample) and a subregion of an entire flat chip (extended sample), respectively. In both cases, a Dectris Eiger detector (514$\times$1030 pixels) was used to acquire diffraction patterns as X-ray beam scanned across the sample, and the 2D diffraction images were saved along the time dimension to form a 3D matrix array (19500$\times$514$\times$1030 for chip pillar and 16800$\times$514$\times$1030 for flat chip). In the data analysis, domain experts usually cropped only central region of the diffraction pattern that contains X-ray signals (lots of zeros outside this region). To fairly assess our compression strategy without giving an overestimated compression ratio, we cropped only central 256$\times$256 pixels.

\vspace{-0.5em}
\subsection{Data characterization and pipeline customization}
\begin{figure}[ht]
\centering
\includegraphics[width=0.8\columnwidth]{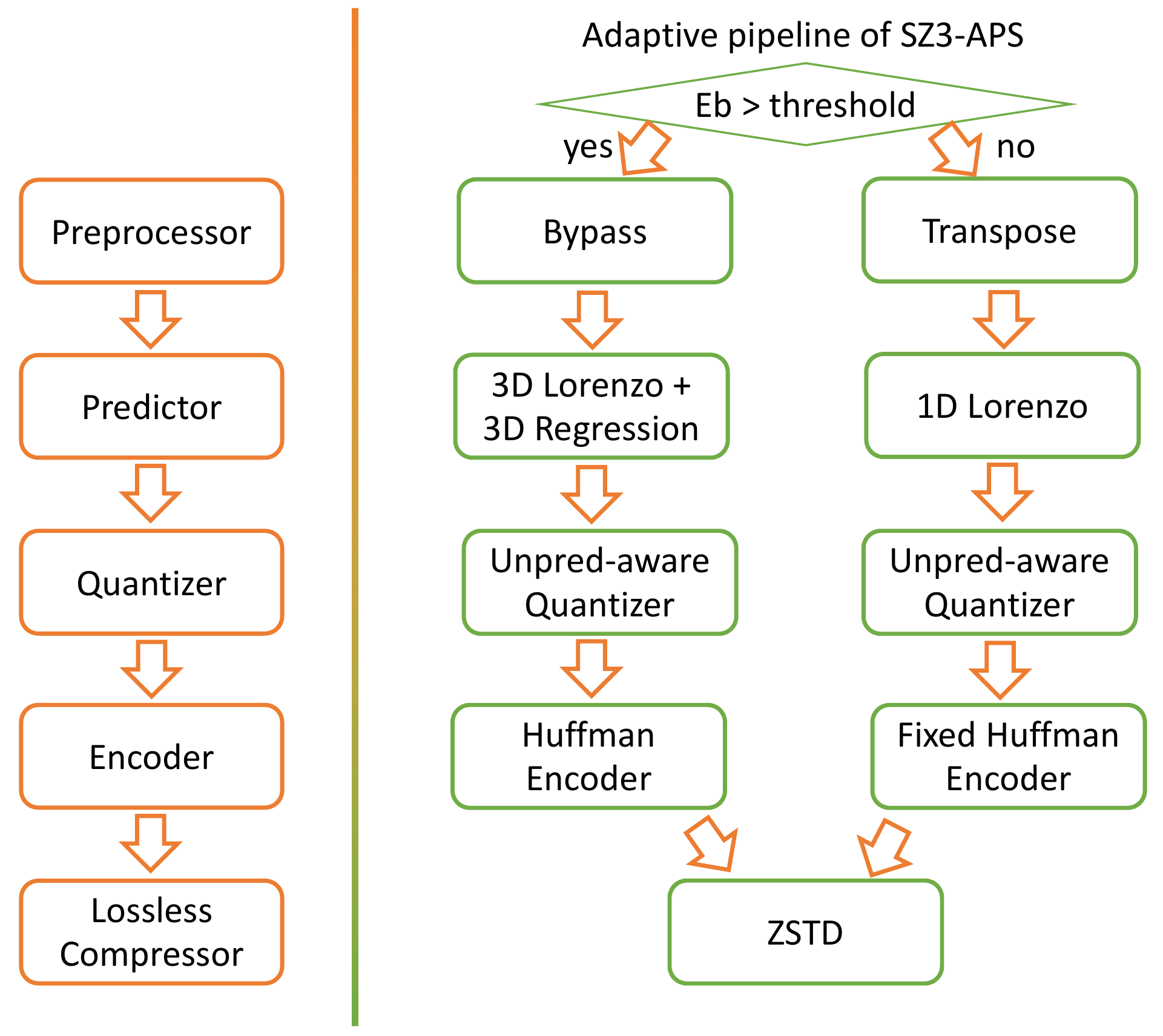}
\vspace{-1em}
\caption{Adaptive compression pipeline for APS data.}
\label{fig:aps-pipeline}
\vspace{-1em}
\end{figure}

We design an adaptive compression pipeline for APS data based on the following analysis. 
First, multidimensional Lorenzo predictor introduces higher noise because more decompressed data values are used for prediction~\cite{sz-reg}, even though it is usually superior to the one-dimensional one by exploiting the multidimensional correlation.
Second, although APS data has three dimensions (e.g., $19500\times256\times256$ for the chip pillar sample), it is actually a stack of 2D images along the time dimension with relatively low spatial correlation. 
When the spatial correlation is not strong, the benefit of using the multidimension Lorenzo predictor may not be able to make up the cost for the higher noise. 
In addition, considering the usually high correlation in time compared with that in spatial region, it might be more effective to compress the data along the time dimension, namely, treating the data as $256\times256$ 1D time series. 
On the other hand, the multidimensional regression-based predictor should be included because it leverages the multidimensional correlation without being affected by the decompression noise~\cite{sz-reg}, which yields good performance when error bound is relatively high. 
This requires switching predictors based on the error bound: using a traditional multialgorithm predictor that involves regression for high error bounds and a customized 1D Lorenzo predictor with a transposition preprocessor that reorganizes the data along the time dimension for low error bounds.
In our implementation, we switch to the latter along with quantization bin width $2$ when the user-specified absolute error bound is less than $0.5$ since this setting generates lossless compression.
Under such circumstance, the noise introduced by using decompressed data is reduced to $0$ when the unpred-aware Quantizer is leveraged, thanks to the restricted quantization bin and the principle of embedded encoding. We further employ a fixed Huffman encoder for fast encoding with comparable compression ratios. 
The corresponding compression pipeline for APS data is depicted in Figure \ref{fig:aps-pipeline}.

\subsection{Evaluation results}
\begin{figure}[ht]
\centering
\subfigure[{chip\_pillar}]
{
\hspace{-3mm}
\raisebox{-1cm}{\includegraphics[width=0.48\columnwidth]{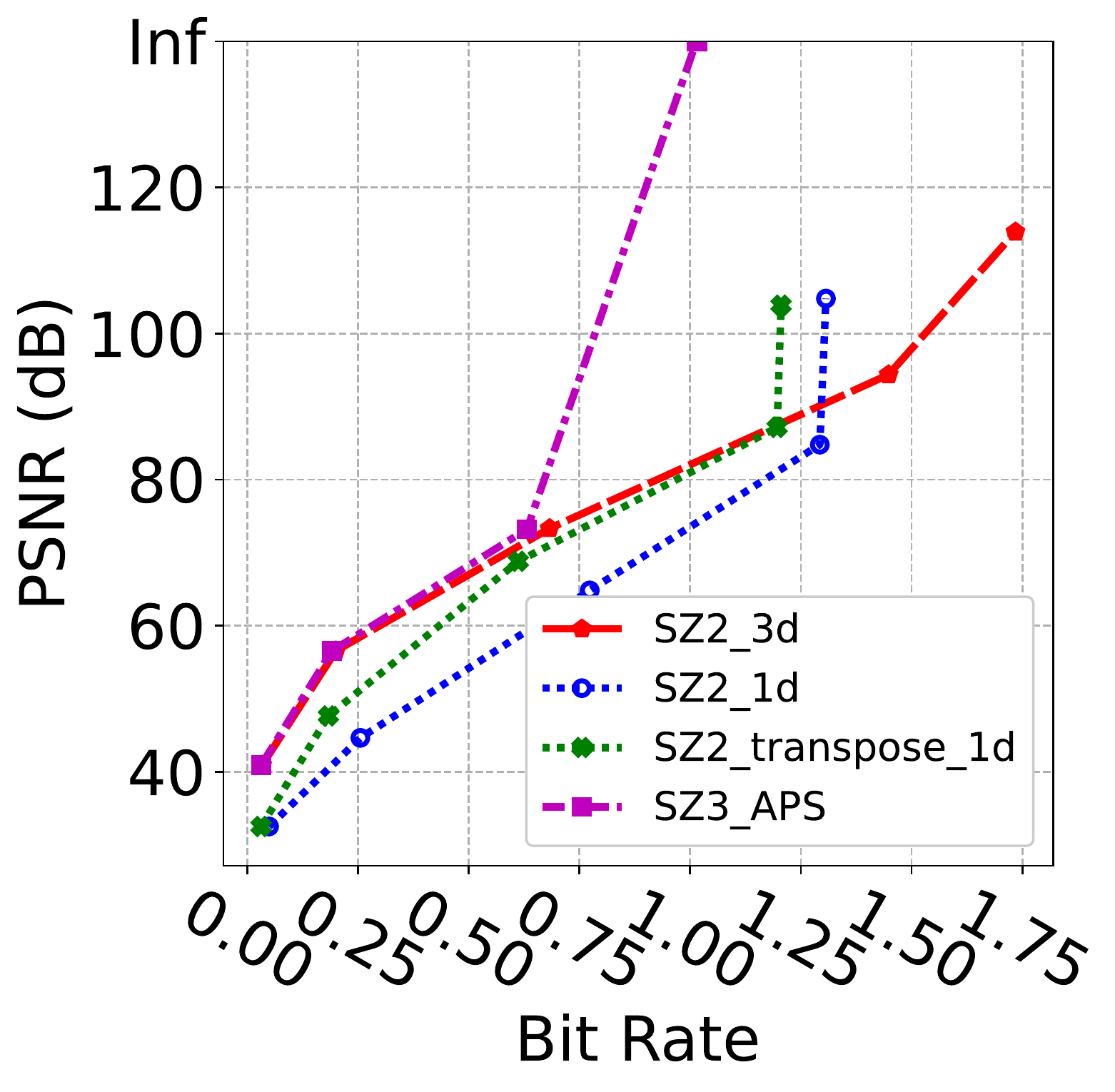}}
}
\subfigure[{flat\_chip}]
{
\hspace{-3mm}
\raisebox{-1cm}{\includegraphics[width=0.48\columnwidth]{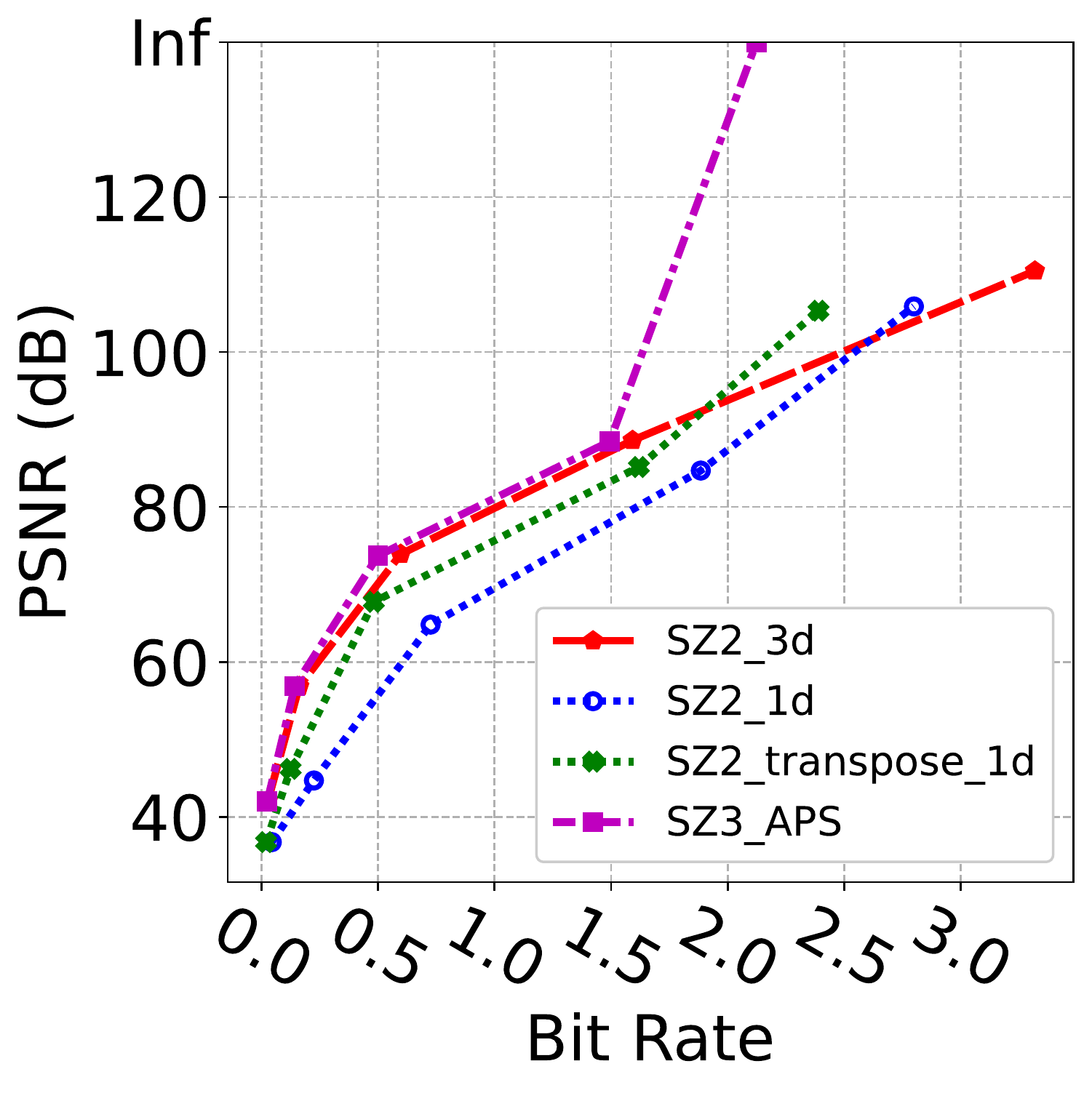}}
}
\vspace{-1em}
\caption{Rate-distortion on APS data.}
\label{fig:aps-rate-distortion}
\vspace{-1em}
\end{figure}

We evaluate the customized APS compressor and compare it with 3 baselines: the generic SZ-2.1 compressor for 1D, 3D, and transposed 1D data.
As illustrated in Figure~\ref{fig:aps-rate-distortion}, a 3D compressor leads to higher PSNR under low bit-rate (high compression ratios), but it suffers when the bit-rate increases to a certain level, where there is a sharp increase in the compression quality for 1D compressors. 
This is caused by the fact that the noise introduced by decompressed data is mitigated with such an error setting in this dataset.  
SZ-2.1 is not aware of this information and incorrectly estimates the Lorenzo prediction noise, leading to the selection of regression predictor even when Lorenzo predictor is better. SZ3-APS adaptively chooses the compression pipeline based on the error bound, which leads to comparable performance to that of SZ-2.1 for 3D data when error bound is high. 
Furthermore, the adopted Unpred-aware Quantizer exhibits higher compression ratios in low error bound, since it provides near-lossless decompressed data that improves the prediction efficiency of the Lorenzo predictor. 
In absolute terms, when the decompression data is near lossless (i.e., error bound less than $0.5$), the compression ratio gain of the proposed compression pipeline is $18\%$ on chip pillar and $12\%$ on flat chip compared with the second best one. 
Note that SZ3-APS turns out to be lossless in this case, which leads to infinity PSNR in the figure. 

%% file: tex/usecases/pipeline.tex
\section{Sustainability, Quality, and Performance Investigation of SZ3}
\label{sec:pipelines}

In this section we first discuss the sustainability of SZ3, and then leverage SZ3 to characterize the quality and performance of diverse compression pipelines.

\subsection{Sustainability}\label{sec:sustainability}
We design SZ3 with modularity in mind to allow for a composable framework with high sustainability. 
Specially, we compare the design of SZ3 with that of SZ2~\cite{sz-repo}, one of the leading error-controlled lossy compressors with prediction-based pipeline, to demonstrate its superiority. 

\subsubsection{The codebase of SZ2}
SZ2 has a large codebase including more than 120 functions with little code reuse, as shown in Table~\ref{tab:sz2-impl}. For example, SZ2 has separate functions to handle the compression or decompression on a dataset with a specific data type, although the logic to compress and decompress different data types is similar. As a result, SZ2 needs to maintain separate code for each data type.  

\begin{table}
\centering
\footnotesize
\caption{SZ2 contains more than 120 functions to support different data types, data dimensions, and data-processing methods}
\label{tab:sz2-impl}
\vspace{-0.5em}
\begin{adjustbox}{width=0.8\columnwidth}
\begin{tabular}{|c|c|}
\hline
\begin{tabular}[c]{@{}c@{}} 
Data \\ Type 
\end{tabular} & \begin{tabular}[c]{@{}c@{}} 
FP32, FP64\\
INT8, INT16, INT32, INT64\\ 
UINT8, UINT16, UINT32, UINT64
\end{tabular}  \\ \hline
\begin{tabular}[c]{@{}c@{}} 
Data \\ Dimension 
\end{tabular} 
& 1D, 2D, 3D, 4D\\\hline
Functionality & \begin{tabular}[c]{@{}c@{}} 
Compression\\
Decompression\\
Parameter Optimization
\end{tabular}  \\ \hline
\end{tabular}
\end{adjustbox}
\vspace{-1em}
\end{table}


The lack of software architecture design makes it difficult and time-consuming to modify and extend the functionality of SZ2. 
With more than 120 functions to update,  some of them are likely to be missed when adding new features to SZ2. Furthermore, the complexity of SZ2 brings challenges to fully validate the correctness of newly added features, because it is time-consuming to write test code that achieves high code coverage for so many functions of SZ2.

\subsubsection{The Codebase of SZ3}
We propose three technologies in SZ3 to improve the code sustainability dramatically, namely compile-time polymorphism, datatype abstraction, and multidimensional iterator. 

\textbf{Compile Time Polymorphism:} 
SZ constructs the composed compression pipelines at compile time, because compile time polymorphism provides an efficient way to switch different implementations of modules to avoid runtime performance downgrade. For implementation, the module instances are placed as the template parameters of the compressor (see Appendix \ref{appd:compress}).
A static assert is executed during the construction of the compressor to ensure that only classes that inherent from specific module interfaces are allowed to be used to initialize template parameters.

\textbf{Datatype Abstraction:} 
We adopt datatype abstraction to simplify the codebase of SZ3 significantly. 
Most module interfaces, implementations, and compressor pipelines in SZ3 are designed with datatypes as template parameters for efficient code reuse. 
By comparison, SZ2 has separate implementations for each datatype, which result in a large code base without code reuse.

\textbf{Multidimensional Iterator:}
A multidimensional iterator is designed in SZ3 to support data access patterns of different dimensions. This is totally different from SZ2, where independent implementations are required for each dimensionality. 
The multidimensional iterator in SZ3 provides a simple API to access the current and nearby data points and move to another position. The boundary situations are handled inside iterators. The iterator design eliminates the need to write separate code based on the data dimensions. The pseudocode of prediction and quantization using the multidimensional iterator is presented in Appendix \ref{appd:pd}.
  
With the multidimensional iterator, the complex nested-loop to iterator through the data and the boundary condition checking  are hidden from the users. The multidimensional iterator also supports arbitrary movement. For example,  to change a 3D iterator to its upper left neighbor, developers can simply use iterator.move(-1, -1, -1) instead of calculating the offset for three dimensions. 

\subsection{Pipeline integration and evaluation}
We integrate three compression pipelines using SZ3 and reveal their suitable cases in terms of quality and performance. Details of the three pipelines are described as follows.

\begin{figure}[ht] 
\centering
\hspace{-10mm}
\subfigure[{RTM}]
{
\raisebox{-1cm}{\includegraphics[scale=0.3]{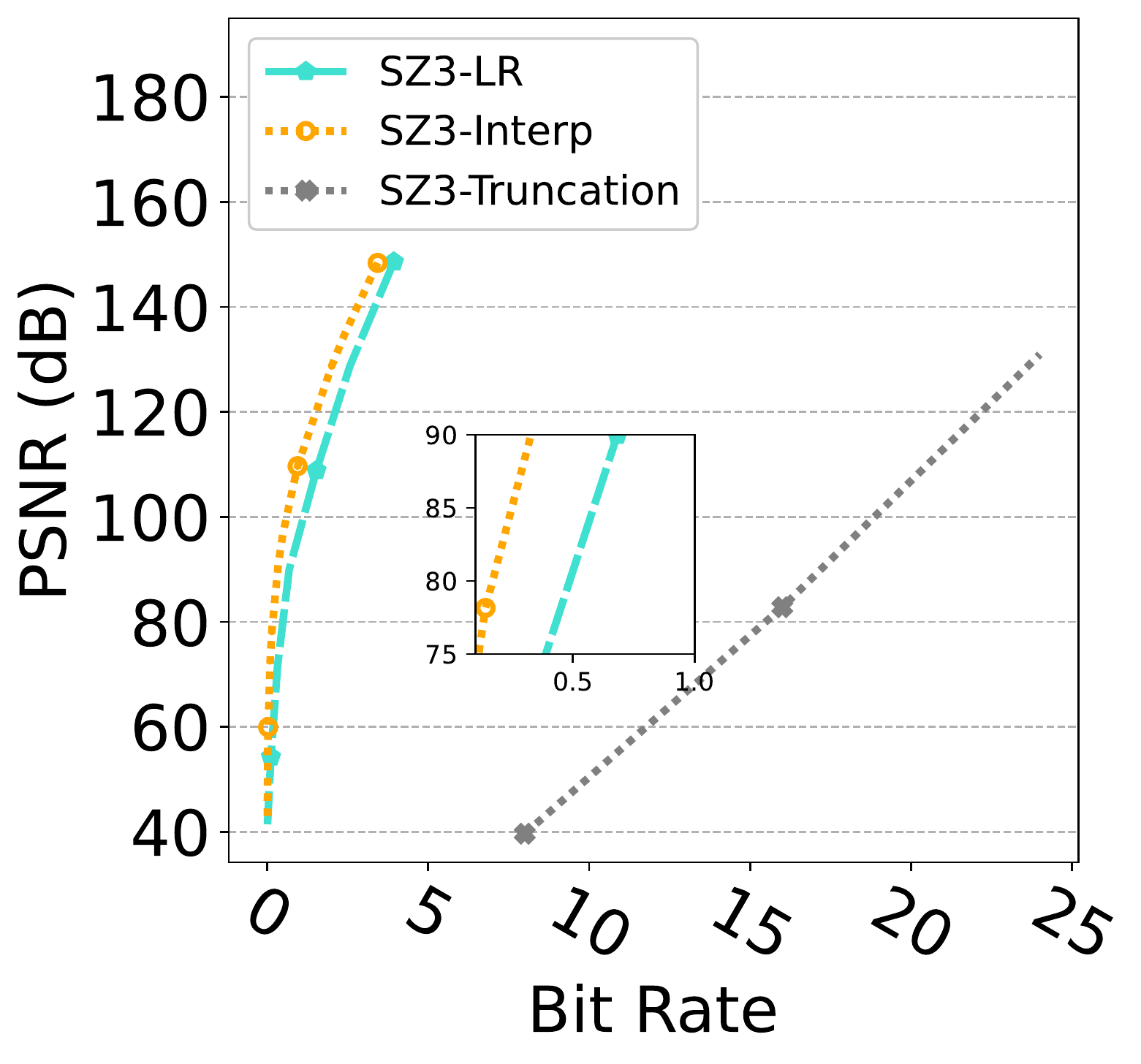}}
}
\hspace{-6.3mm}
\subfigure[{NYX}]
{
\raisebox{-1cm}{\includegraphics[scale=0.3]{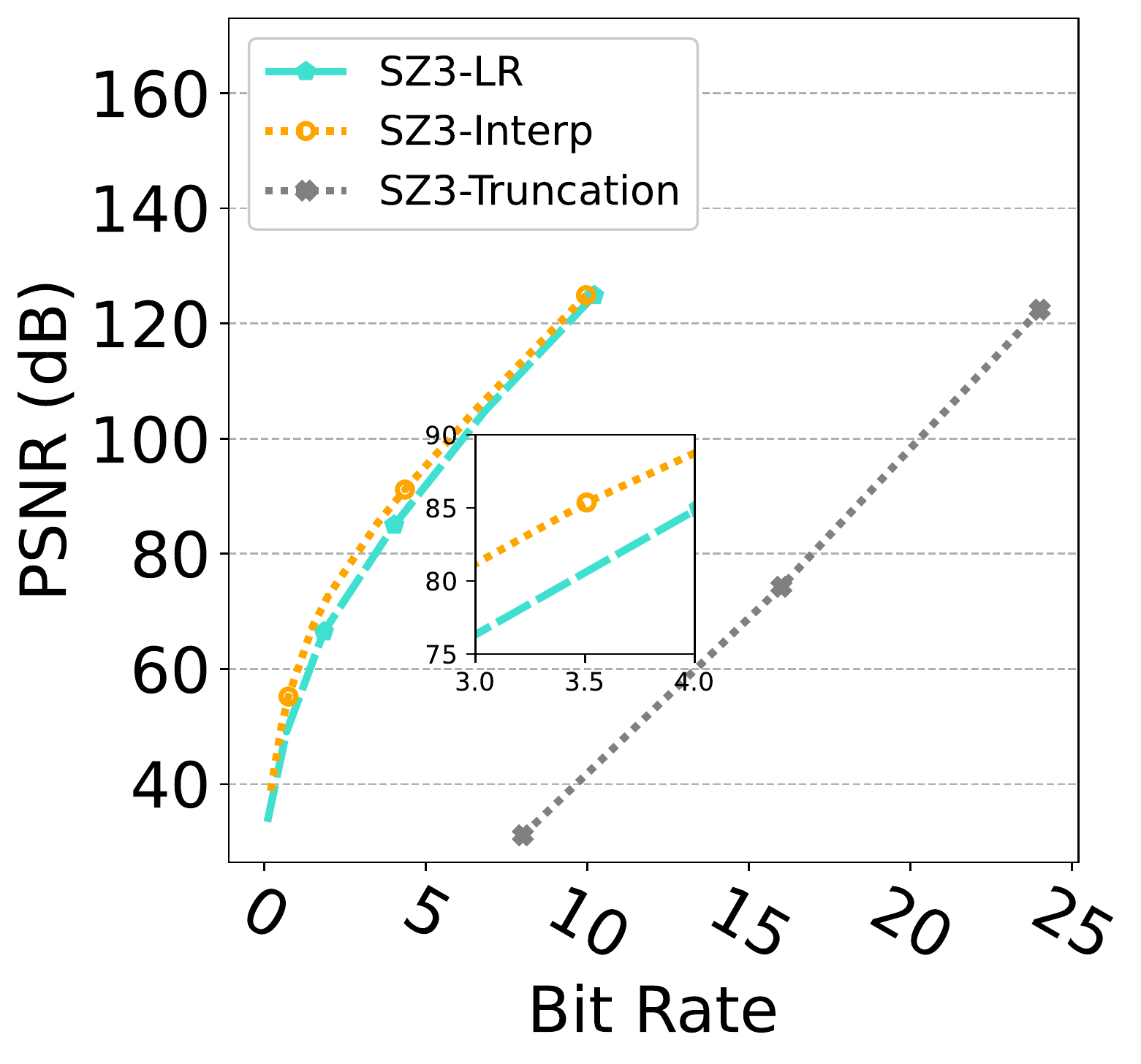}}
}
\hspace{-9mm}

\hspace{-9mm}
\subfigure[{Miranda}]
{
\raisebox{-1cm}{\includegraphics[scale=0.30]{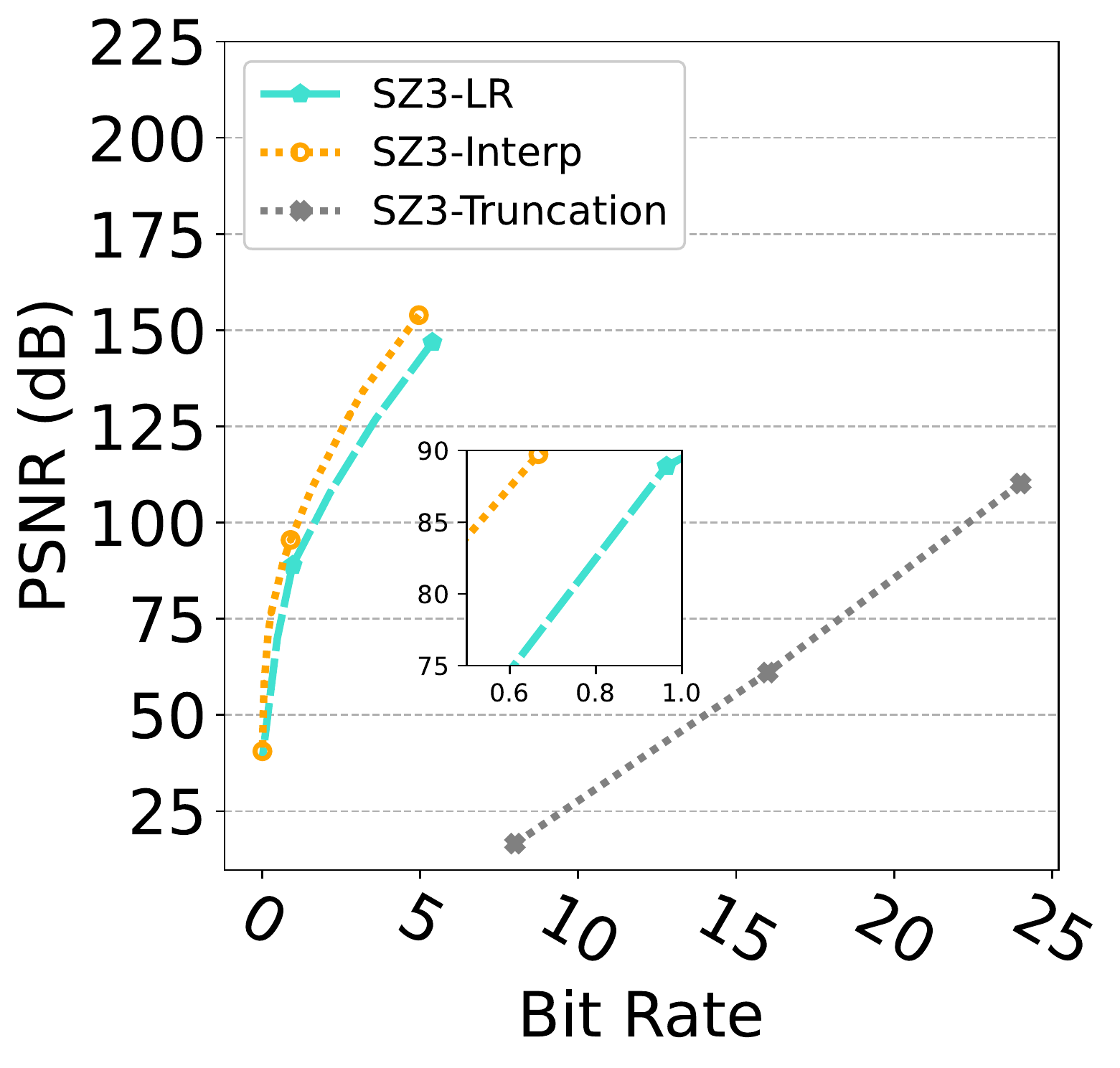}}
}
\hspace{-5mm}
\subfigure[{Scale-LETKF}]
{
\raisebox{-1cm}{\includegraphics[scale=0.30]{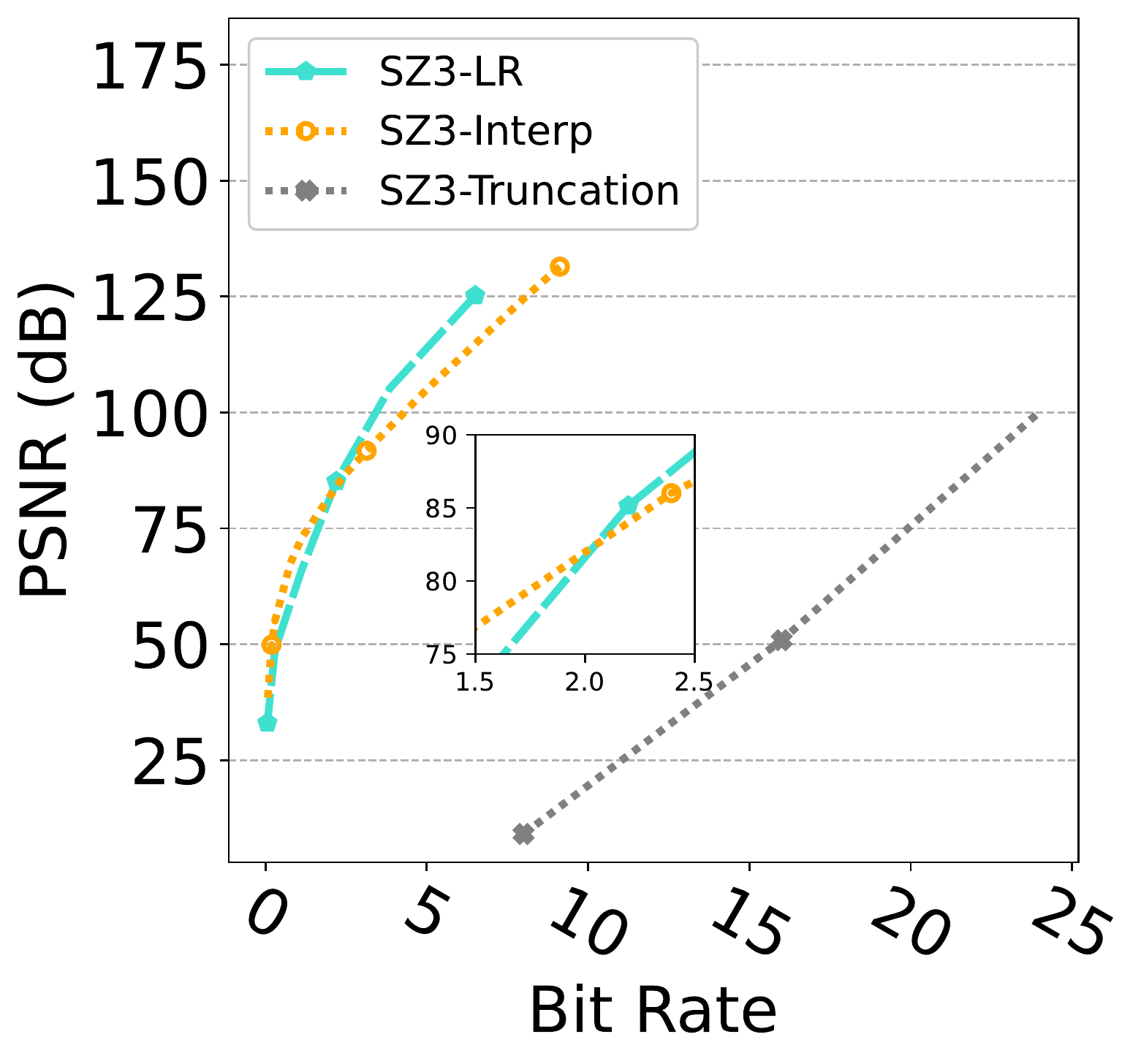}}
}
\hspace{-10mm}

\hspace{-8mm}
\subfigure[{QMCPack}]
{
\raisebox{-1cm}{\includegraphics[scale=0.30]{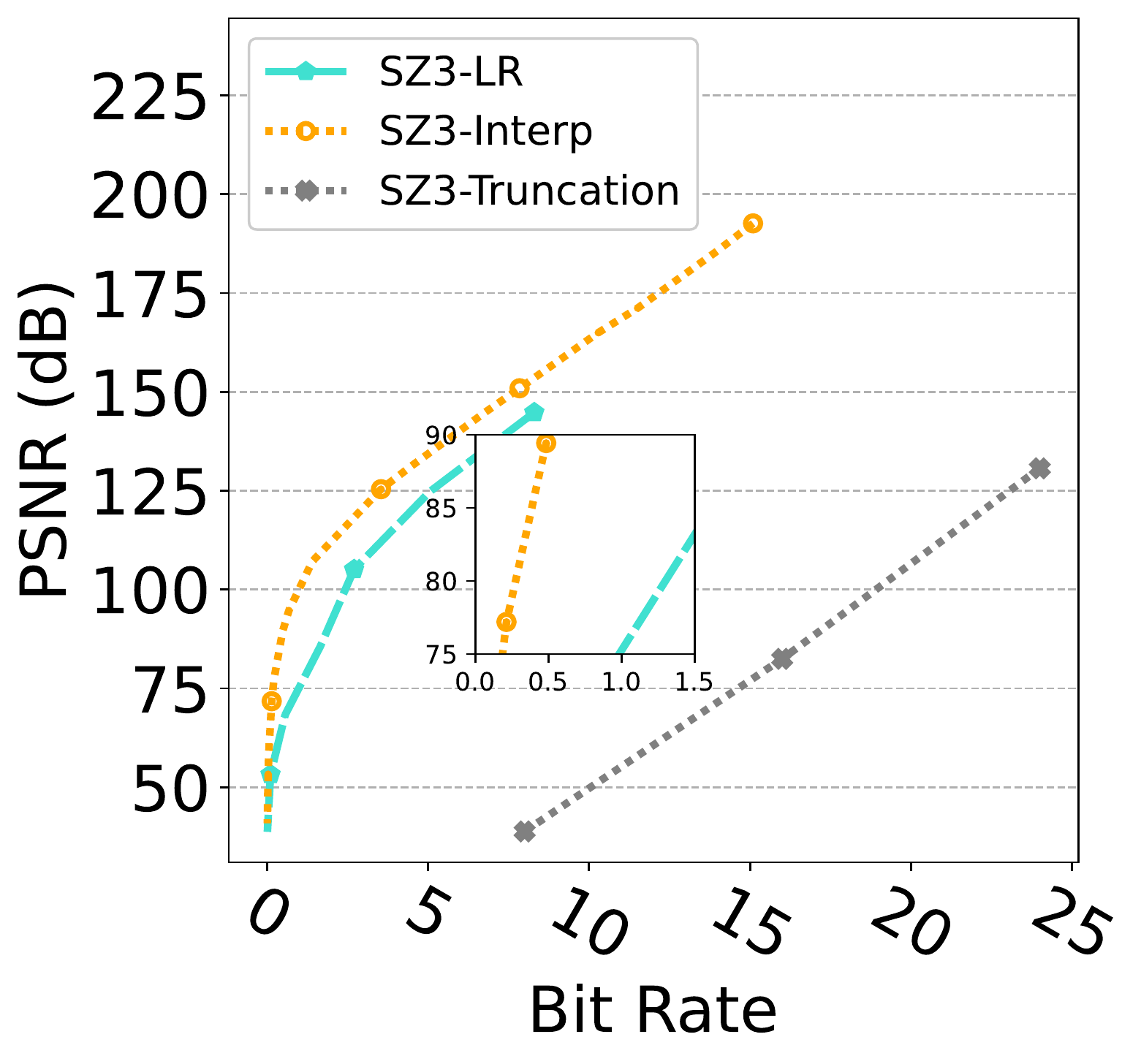}}
}
\hspace{-5mm}
\subfigure[{Hurricane}]
{
\raisebox{-1cm}{\includegraphics[scale=0.30]{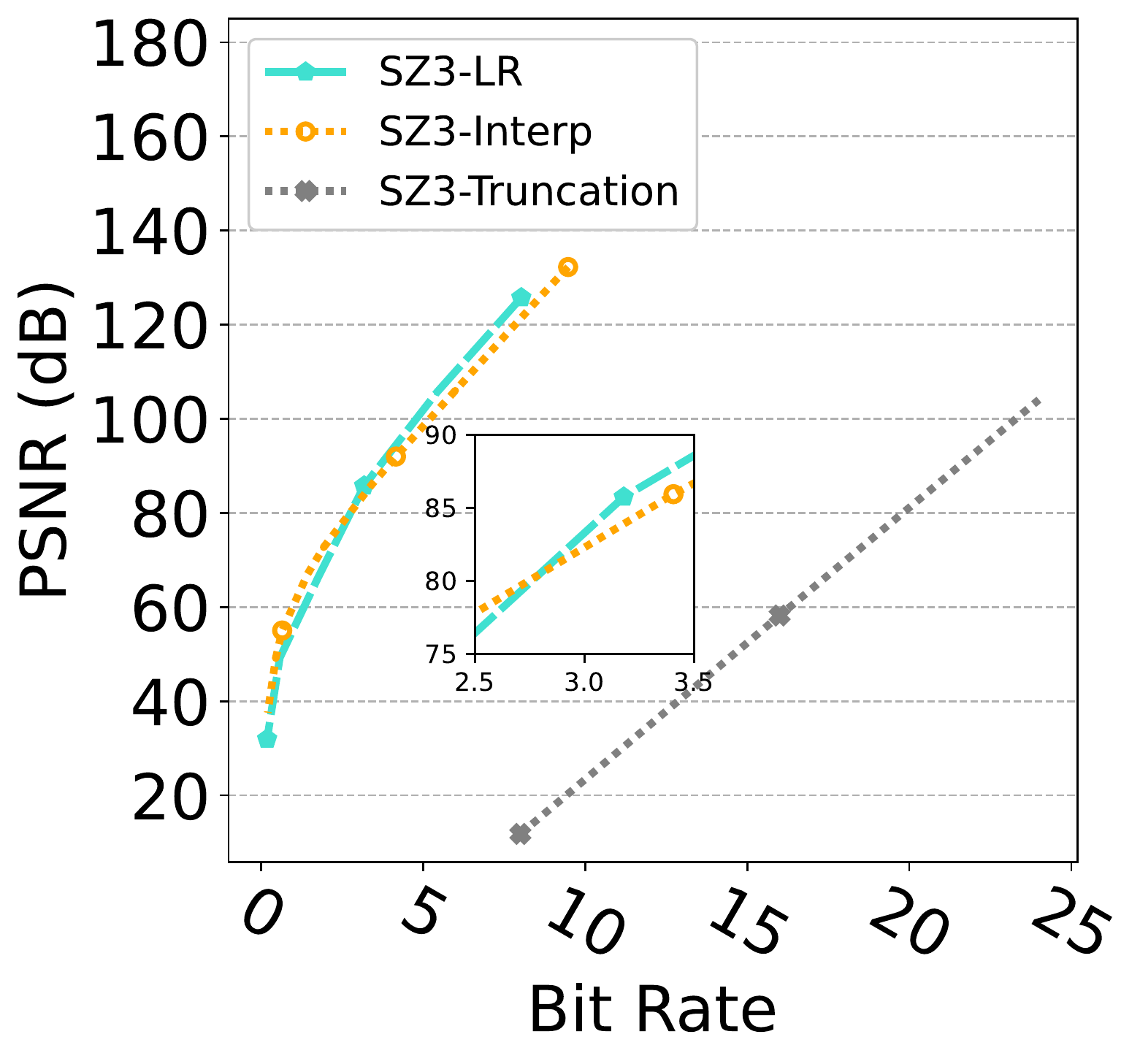}}
}
\hspace{-7mm}
\vspace{-0.5em}
\caption{Compression quality evaluation (lower bit rate \& higher PSNR $\rightarrow$ better quality). Result for SZ2.1 is omitted since it is very similar to that of SZ3-LR.}
\label{fig:rate-distortion}
\vspace{-1em}
\end{figure}

\begin{figure}[ht] \centering
\hspace{-8mm}
\subfigure[{Compression}]
{
\raisebox{-1cm}{\includegraphics[scale=0.27]{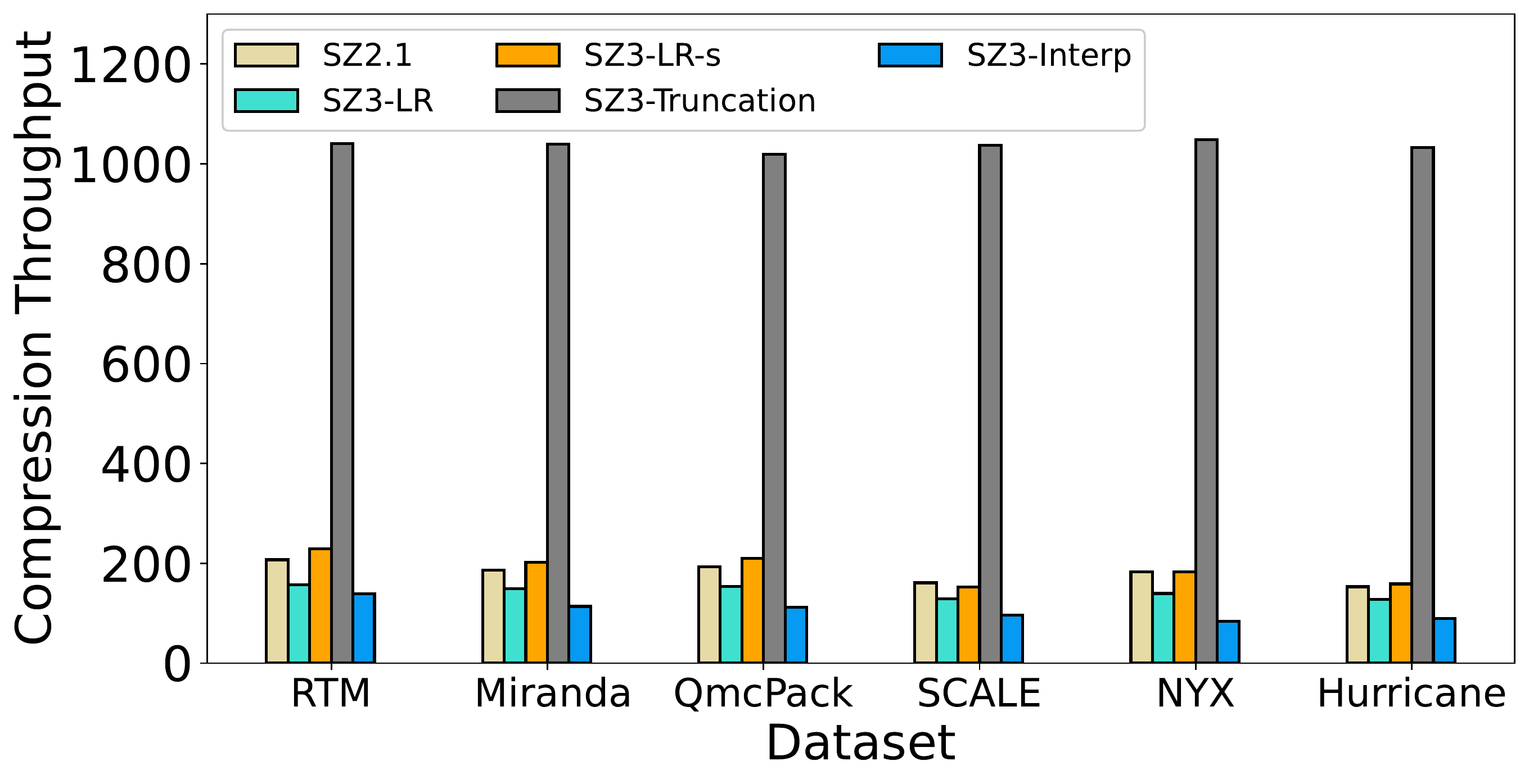}}}

\hspace{-8mm}
\subfigure[{Decompression}]
{
\raisebox{-1cm}{\includegraphics[scale=0.27]{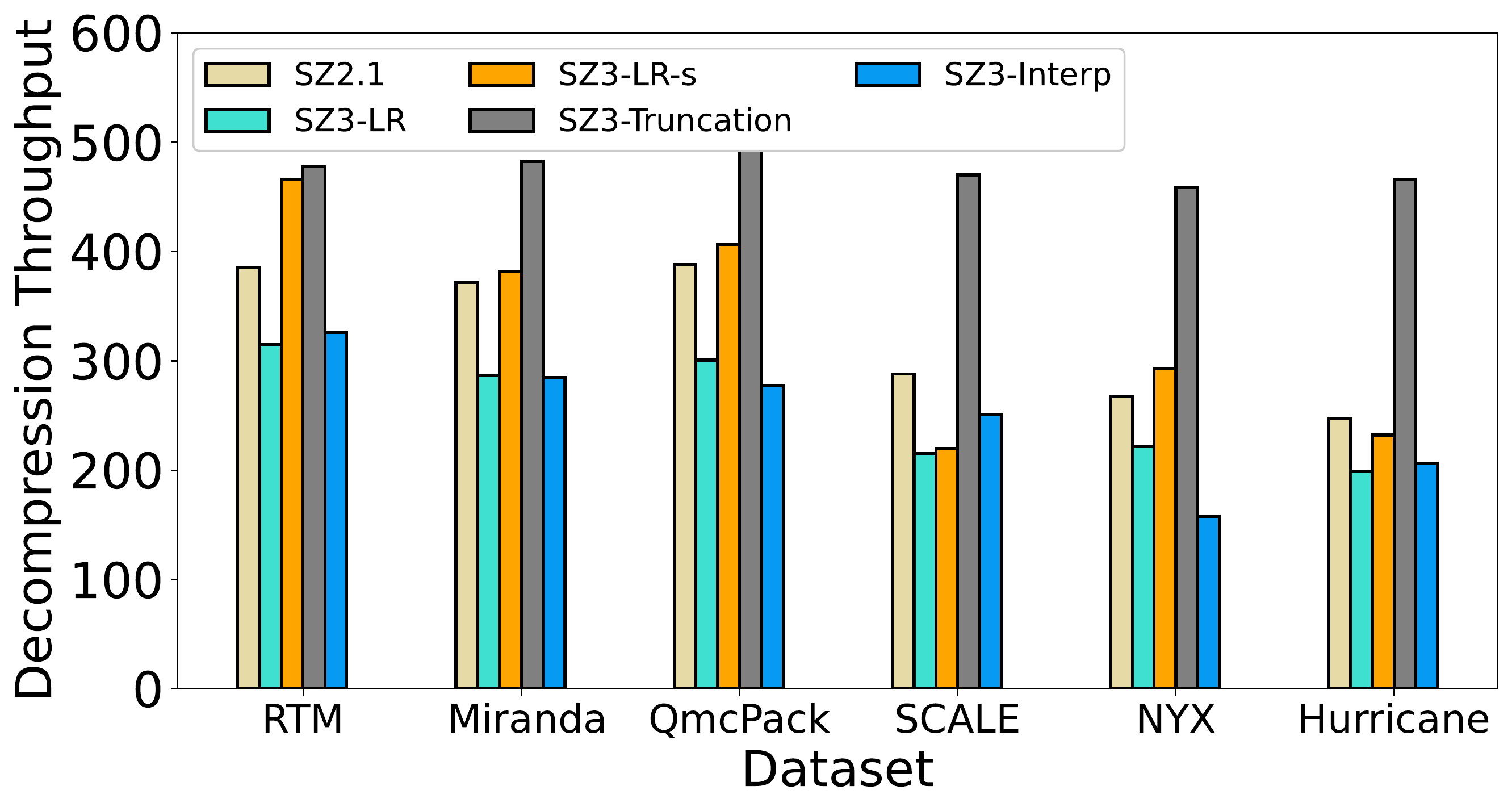}}}

\vspace{-1em}
\caption{Compression/decompression throughput (MB/s) when relative error bound (error bound normalized to value range) is 1E-3.}
\label{fig:speed}
\vspace{-1em}
\end{figure}





\textbf{Compression Pipeline SZ3-LR:}~
SZ3-LR is the implementation of the classic compressor SZ2~\cite{sz-reg} using SZ3's modular mechanism, which relies on a multialgorithm predictor for better data correlation. This predictor consists of a Lorenzo predictor and a regression-based predictor and predicts data using the better result  in between based on blockwise error estimation. 
As depicted in Figure~\ref{fig:sz3-arch}, it uses a linear-scaling quantizer and a Huffman encoder and the \emph{zstd} lossless compressor in the other stages. 

\textbf{Compression Pipeline SZ3-Truncation:}~
SZ3-Truncation is a very fast compression pipeline designed for cases where speed is more important than compression ratio. Given the target bytes $k$ as input parameter, it keeps $k$ most-significant bytes of each floating-point data while discarding the rest of the  bytes. 
To achieve high compression speed, it bypasses the other stages, which in turn leads to low compression ratios in general cases.  

\textbf{Compression Pipeline SZ3-Interp:}~
SZ3-Interp has interpolation-based predictors~\cite{zhao2021optimizing} in its pipeline. Both linear interpolation and cubic spline interpolation are included, and they are better than Lorenzo and regression predictors in many cases for the following reasons. On the one hand, interpolation-based predictors are not affected by the error accumulation effect that is normal in Lorenzo predictor, because the predicted value is based on previous data points in the Lorenzo predictor while it is based on coefficients in interpolation-based predictors. On the other hand, unlike linear regression, which has an overhead to store coefficients, SZ3-Interp has constant coefficients and therefore does not have storage overhead. 
Similar to SZ3-LR, it uses a linear-scaling quantizer for respecting error bounds, as well as a Huffman encoder and the \emph{zstd} lossless compressor for high compression ratios. 

We use datasets from five scientific domains :  cosmology, climate, quantum structure, seismic wave, and turbulence. The detailed information is shown in Table \ref{tab:data information}.

\begin{table}[ht]
\centering
\caption{Dataset Information}
\label{tab:data information}
\vspace{-2mm}
\resizebox{\linewidth}{!}{
\begin{tabular}{|c|c|c|c|c|}
\hline
 Application & Domain & \#Fields & Dimensions & Total Size\\
\hline
HACC & Cosmology & 6 & $280 \times 953 \times 867$ & 6.3GB\\
\hline
ATM & Climate &77 & $1800 \times 3600 $& 1.9GB\\
\hline
Hurricane  & Climate &13& $100 \times 500 \times 500$ & 1.2GB\\
\hline
NYX  & Cosmology &6& $512 \times 512 \times 512$ & 3GB\\
\hline
SCALE-LETKF & Climate & 6 & $98 \times 1200 \times 1200 $ & 3.2GB\\
\hline
QMCPack & Quantum Structure & 1& $288 \times 115 \times 69 \times 69$ & 0.6GB \\
\hline
RTM & Seismic Wave & 3600& $449 \times 449 \times 235$ & 635GB \\ \hline
Miranda & Turbulence & 7 & $256 \times 384 \times 384 $& 1GB \\ \hline
\end{tabular}
}
\vspace{-1em}
\end{table}

We demonstrate the compression quality of the three pipelines using rate-distortion graph in Figure~\ref{fig:rate-distortion}. 
Note that the rate distortion of SZ2.1 is identical to that of SZ3-LR; thus we do not show SZ2.1 in this figure. We observe from Figure~\ref{fig:rate-distortion} that SZ3-Truncation has the lowest compression quality, and this is consistent with its simple byte-truncation design. SZ3-Interp is better than SZ3-LR on most of the datasets, especially on cases with a high compression ratio  with a  bit rate lower than 3. For example, on the Miranda dataset, under the same PSNR of 90, the compression ratio of SZ3-Interp is 47, and it is 56\% higher than the  compression ratio of SZ3-LR, which is 30. On the other hand, SZ3-LR is still the best choice on the Scale and Hurricane datasets when high compression accuracy is needed. 

The performance evaluation is shown in Figure~\ref{fig:speed}. We include SZ2.1 as the baseline. SZ3-LR-s) is a performance-oriented version of SZ3-LR that shares the same logic but has a different implementation of the predictor module with SZ3-LR. The predictor module in SZ3-LR uses  a multidimensional iterator for better code simplicity, and in SZ3-LR-s the predictor contains several codecs, each of which handles data in a specific dimension. Note that SZ3-LR-s still has a modular design and can be customized with different a quantizer, encoder, and lossless compressors. We can see from Figure~\ref{fig:speed} that SZ3-LR-s has comparable performance with SZ2.1 on all datasets. SZ3-Truncation has the best performance among all compressors including SZ2.1. Its 1GB/s compression throughput is 4X higher than that of  the second-best compressor. SZ3-Interp is not as fast as others, but its throughput is still higher than 100  MB/s in all cases. 

The quality and performance evaluations reveal the suitable cases for the three built-in pipelines. Specifically, SZ3-Trunction, as a high-speed compressor, is the best choice when there are strict requirements on the compression time, as with some in situ applications. SZ3-Interp would be the first preference in cases where high compression ratio is wanted under relaxed time constraints, such as scientific applications that run for a long time and generate large amounts of data. SZ3-LR has balanced quality and speed; users could choose it as the default compressor in general situations where both high compression ratio and short compression time are needed.

%% file: tex/conclusion.tex
\section{Conclusion and Future Work}
\label{sec:conclusion}
In this paper, we propose a modular, composable compression framework --SZ3-- which allows users to customize on-demand error-bounded lossy compressors in an adaptive and extensible fashion with minimal effort.  
Using SZ3, we develop efficient error-bounded lossy compressors for two real-word application datasets based on the data characteristic and user requirements, which improve the compression ratios by 20\% when compared with other state-of-the-art compressors with the same data distortion. 
We also compare the sustainability of SZ3 with existing compressors, and leverage it to integrate and evaluate different compression pipelines.
In the future, we will integrate more instances to the framework for diverse use cases and provide support for various hardware including GPUs and FPGAs.